\documentclass[linenumbers]{emulateapj}
\usepackage{graphicx} 
\usepackage{lipsum}
\usepackage{graphicx}
\usepackage{hyperref}

\usepackage{amsmath}
\usepackage{longtable}
\def\dbar{{\mathchar'26\mkern-12mu d}}

\shorttitle{Time Dependent Photoionization}
\shortauthors{Sadaula et al.}

\graphicspath{{./}{figures/}}
\begin{document}
\title{Time Dependent Photoionization Modeling of Warm Absorbers in Active Galactic Nuclei}

\author{Dev R Sadaula}
\affiliation{Western Michigan University \\
1903 W Michigan Ave,\\
Kalamazoo, MI 49008, USA}
\affiliation{NASA Goddard Space Flight Center \\
8800 Greenbelt Rd,\\
Greenbelt, MD 20771, USA}

\author{Manuel A Bautista}
\affiliation{Western Michigan University \\
1903 W Michigan Ave,\\
Kalamazoo, MI 49008, USA}

\author{Javier A Garc{\'\i}a}
\affiliation{California Institute of Technology \\
1200 E California Blvd,\\
 Pasadena, CA 91125, USA}
\affiliation{ Dr. Karl Remeis-Observatory and Erlangen Centre for Astroparticle Physics\\
Sternwartstr. 7, \\
D-96049 Bamberg, Germany}

\author{Timothy R Kallman}
\affiliation{NASA Goddard Space Flight Center \\
8800 Greenbelt Rd,\\
Greenbelt, MD 20771, USA}

\begin{abstract}

Warm absorber spectra contain bound-bound and bound-free absorption features seen in the X-ray and UV spectra from many active galactic nuclei (AGN).  The widths and centroid energies of these features indicate they occur in outflowing gas, and the outflow can affect the gas within the host galaxy.  Thus the warm absorber mass and energy budgets are of great interest.  Estimates for these properties depend on models which connect the observed strengths of the absorption features with the density, composition, and ionization state of the absorbing gas.  Such models assume that the ionization and heating of the gas come primarily from the strong continuum near the central black hole.  They also assume that the various heating, cooling, ionization, and recombination processes are in a time-steady balance.  This assumption may not be valid, owing to the intrinsic time-variability of the illuminating continuum or other factors which change the cloud environment.  This paper presents models for warm absorbers which follow the time dependence of the ionization, temperature, and radiation field in warm absorber gas clouds in response to a changing continuum illumination.   We show that the effects of time variability are important over a range of parameter values, that time dependent models differ from equilibrium models in meaningful ways, and that these effects should be included in models which derive properties of warm absorber outflows.

\end{abstract}


\section{Introduction} \label{sec:intro}
Spectroscopy is a powerful tool for studying astrophysical objects. Composition, ionization and excitation state, and motion of the constituent gas can all be found from spectral observations. An important example is active galactic nuclei (AGN) warm absorbers, which show a myriad of blue-shifted absorption lines and edges in the UV and X-ray bands \citep{hol07,hol10}. Blue shifted absorption lines in the spectra reveal the presence of outflows of ionized gas at speed $\sim$1000 km s$^{-1}$ \citep {kas01}. Outflow dynamics may be driven by thermal expansion, radiation pressure, or magneto-hydrodynamic effect \citep{cre03}. Models for warm absorbers have reached a level of sophistication that permits very detailed comparison with observations. However, the fundamental properties of warm absorbers are still not fully understood. Progress towards understanding warm absorbers is limited by a key assumption used in modeling these flows: that the gas is in ionization and excitation equilibrium. This is likely not correct in detail; the dynamical timescales in the flow and the intrinsic variability timescale of the AGN may be comparable to the timescales characterizing the ionization and excitation and the inverse processes in the gas responsible for the lines we observe. This means that the gas departs from equilibrium, so temporal dependency for ionization, heating, cooling, and radiative transfer should be included in the modeling of the absorber. In this paper, we explore the inclusion of time dependence in photoionization calculations.

Warm absorber (WA) features in X-ray spectra were first recognized by \cite{hal84} in the Seyfert galaxy MR 2251-178 observed by the EINSTEIN Satellite. It is now known that the spectra of many AGN show deep absorption features in the spectral range from 0.7 to 1.5 keV \citep{rey97}.  It is believed that this gas is ionized and heated mainly by photoionization by the intense radiation field from the central compact object. Warm absorbers are seen in approximately half \citep{roz05} of Seyfert galaxies. This suggests that warm absorber gas covers a substantial part of the central ionizing source. Most absorption lines are blueshifted with respect to the galactic nucleus by $\sim 10^3$ km s$^{-1}$ \citep{kas02}.  Some observations reveal ultrafast outflows with much greater apparent outflow speed \citep{cha02,pau103,pau203}.

Photoionization is of widespread importance in astrophysics. Strong continuum radiation  illuminates gas and is absorbed, reprocessed, and re-emitted. This is the dominant process of heating in  photoionized plasmas. Available code packages for modeling photoionized plasmas include  {{\sc XSTAR}} \citep{kall01}, Cloudy \citep{fer98}, SPEX \citep{kaa96} and MOCASSIN \citep{erc03}.  It is common for photoionization models to assume the gas is in an equilibrium state.  This is used to simultaneously solve the ionization balance, thermal balance, and radiative transfer equation. If so, the most important free parameters are density, spectral energy distribution, flux, column density, and elemental abundances. These can be adjusted repeatedly while performing simulation calculations in order to fit observations. 

Equilibrium calculations derive typical total hydrogen column densities ($N_H$) of WA $\sim 10^{22}$ cm$^{-2}$, temperatures of $\sim~10^5$ K, and total hydrogen number densities ($n_H$) $\sim 10^7$ cm$^{-3}$ (hereafter, column density means total hydrogen column density and gas density means total hydrogen number density unless otherwise specified).
To a good approximation, the ionization state of photoionized gas can be parameterized by the ionization parameter $\xi$ and defined as,
\begin{equation}
    \xi = \frac{L_{ion}}{n_HR^2} = \frac{4\pi F_{ion}}{n_H}    
\end{equation}
where $L_{ion}$ is the ionizing luminosity of the source, $F_{ion}$ is the ionizing flux, and both of these quantities are integrated from 1-1000~Ry. $R$ is the distance of gas from the ionizing source. This definition is widespread in X-ray problems, and it was first introduced by \cite{tarter69}. The other way of defining ionization parameter is $U=\int_{\varepsilon_{1}}^{\varepsilon_2} L_{\varepsilon}  / (\varepsilon 4 \pi R^2 n_H c) d\varepsilon,$ where, $L_\nu$ is the ionizing luminosity per frequency interval and $c$ is the speed of light. $U$ and $\xi$ are easily convertible to each other for a given spectral energy distribution of the ionizing continuum.  For the spectrum shape that we adopt here, i.e. a single power law from 1 - 1000 Ry with energy index -1, the conversion is $U=\xi/56.64$ erg cm s$^{-1}$. The higher the ionization parameter, the more ionized the gas and vice-versa. Warm absorber components generally exist either in the low ionization state of $ \xi\sim 1$ erg cm $s^{-1}$ or high ionization state of $\xi \sim 100$ erg cm $s^{-1}$ \citep{laha14}.

As a result of the degeneracy between the gas density and the location in the ionization parameter definition, these quantities are not independently well-constrained. Possible WA distances range from the region of the accretion disc, which is at $\sim 0.01$ pc from the AGN center to the dusty molecular torus $\sim 30$ pc. Possible proposed locations include: the accretion disc \citep{elv00,kro07}, the broad-line region \citep{kra05}, the obscuring torus \citep{krolik2001,blu05}, and the narrow-line region \citep{beh03,cre09}. Similar possibilities have been proposed in simulations, such as an accretion disc wind \citep{pro04,ris10}, a wind from torus \citep{dor08}, and large-scale outflows \citep{kur09}.

The equilibrium assumption is valid only when the equilibration timescale for the microscopic processes like excitation, ionization, and thermal balance is much shorter than the timescale of the variation of the ionizing source or the timescale of change in the location or density of the plasma. If the ionizing flux changes at a rate faster than the equilibration timescale or if the conditions in the plasma change on a timescale shorter than the microscopic timescale, then the calculation of the time dependent effects of photoionization is needed. AGN variability has been extensively observed and reveals significant variability ($\geq 20 \% $) on timescales as short as $\sim 10^3$ s.  \citep{sil16}.  Time dependent photoionization has been explored previously in models for  the interstellar medium \citep{lyu96,jou98}, H II regions \citep{rod98, ric00}, planetary nebulae \citep{har76,har77,sch87,fra94, mar97}, novae and supernovae \citep{hau92, bec95, koz98}, reionization of the intergalactic medium \citep{ike86,sha87,fer96,gir96}, ionization of the solar chromosphere \citep{car02}, gamma-ray bursts \citep{per98,bot99}, accretion disks \citep{woo96}, active galactic nuclei \citep{nic97,kro07}, the evolution of the early universe \citep{sea11}, and quasar FeLoBALs \citep{bau10}.  \cite{gna17} calculated the time dependent cooling in photoionized plasma in the context of intergalactic gas. \cite{bau18} modeled the nebula heated by short-period binary stars to describe temperature fluctuations. \cite{mat12}  solved the time dependent radiative transfer equation as applied to  AGN outflows and gamma-ray bursts. \cite{kro07} discussed time dependent modeling as a tool to constrain warm absorber properties.  Most notably, for our purposes, \cite{gar13} calculated the time evolution of H II regions, including the time dependence of radiation transfer.  This serves as the departure point for our work.

In this paper, we  explore time dependent photoionization models applied to AGN warm absorbers. Our calculation differs from many of the previous works on this topic because we solve simultaneously all three coupled time dependent equations affecting the gas: level populations, energy balance or temperature,  and  radiative transfer simultaneously. To carry this out, we upgraded the  photoionization code {\sc xstar} \citep{kall01}. {\sc xstar} is a publicly available command-based computer application that calculates the physical properties of photoionized gases and their emission spectra based on the steady-state approximation. It can be used in a wide range of astronomical applications simply with a proper choice of input parameters. {\sc xstar}  assumes a spherical gas shell surrounds a central source of ionizing radiation. The incident electromagnetic energy is either absorbed or scattered by the gas and reradiated in other parts of the spectrum.  {\sc xstar} calculates such emission and absorption spectra. Additional heat sources (or sinks) such as mechanical compression or expansion or cosmic ray scattering can also be incorporated. The user provides the model parameters such as shape and strength of incident continuum, elemental abundances, density or pressure, and cloud column density; the code returns the ionization structure and temperature, opacity, emitted line, and continuum fluxes. Transport of incident radiation into the cloud, computation of temperature, ionization, and atomic level populations at each place in the cloud, and transfer of emitted radiation out of the cloud are the essential elements in modeling the photoionized gas in {\sc xstar}. We perform time dependent calculations by taking the initial condition from the equilibrium calculation and then calculating the response of the gas properties to a time-dependent change in the illuminating flux.

In the rest of this paper, we  describe the general properties of our time dependent photoionization calculation, and we illustrate with examples taken from the study of the H II regions and warm absorbers. Theory and equations are presented in section 2. Modeling of the warm absorber is discussed in section 3. Results, discussion, and conclusion are presented in sections 4, 5, and 6, respectively.  The numerical method we employ is presented in appendix \ref{if0}.

\section{Theory}\label{sec2}
Our models consist of the solution of the coupled equations describing level populations, temperature, and radiation field. The foundations of our treatment are inspired by and follow the work of \cite{gar13}. Some of the material presented here is a repetition of that paper;   we  repeat it here for completeness, with modifications, and to facilitate our later discussion.

\subsection{Level Populations}

The population of an atomic level obeys the kinetic equation involving the atomic rates into and out of the level:  

\begin{equation}\label{levpop}
    \frac{dn_{i,X}}{dt}=\sum_{j=1}^p n_{j,X} R_{ji}-\sum_{k=1}^p n_{i,X} R_{ik}
\end{equation}
and the equation of number conservation
\begin{equation}\label{chargebalance}
    \sum_{i=1}^p n_{i,X} = x n_H   
\end{equation}
where $n_{i,X}$ are the level populations in $i^{th}$ energy level of element X in units  cm$^{-3}$, which is equal to the product of the fractional elemental abundance and total hydrogen number gas density, $n_H$. $ x$ is the fractional elemental abundance of species $X$ relative to hydrogen. $R_{ji}$ is the transition rate from the $j^{th}$ to $i^{th}$ energy level contributed by atomic processes, including photoexcitation, photoionization, collisional excitation, collisional ionization, recombination, charge transfer, and radiative decay. The value of rates for $R_{ij}$ are taken from XSTAR. $p$ is the total number of energy levels considered for the particular element. As discussed in section 3, we adopt 
a simplified level structure so that, for an element with atomic number Z, p=$2Z+1$. Here and in what follows, we adopt the values of the atomic constants (e.g., photoionization cross-sections, atomic energy levels, collision rate coefficients) from the {\sc xstar} database \citep{bau01}.
\subsection{Energy Balance}
The electron temperature is determined via the first law of thermodynamics. For a closed system,
\begin{equation}
    \dbar Q=dU+PdV
\end{equation}
where $\dbar Q$ is the differential net heat transfer energy density into or out of the gas, the amount of energy density put in the gas, $dU$ is the change in internal energy, and $PdV$ is the differential external work done on or by the gas. Here, all three terms have a unit of the erg. In the absence of external work done by the gas, all the energy put in the gas is used to increase the internal energy. In this case, the above equation reduces to:
\begin{equation}
    \dbar q=du
\end{equation}
here, $q$ and $u$ are defined as differential net heat transfer to the gas and change in the internal energy of the gas in the unit of erg cm$^{-3}$.


The rate of change in the amount of internal energy density stored in the gas is the difference between heating and cooling rates and is given by: 
\begin{equation}
\frac{du}{dt}= \frac{\dbar q}{dt}=\Gamma^{heat}-\Lambda^{cool}
\end{equation}
where $\Gamma^{heat}$ and $\Lambda^{cool}$ are total heating and cooling rates in units of ${\rm erg}\ {\rm s}^{-1}\ {\rm cm}^{-3} $. Assuming classical ideal gas, the internal energy density can be written as $u~=~\frac{3}{2}n_tkT$, where $n_t$ is the total number density of particles, i.e., sum over all electrons, ions, and neutral atoms. With the value of $u$, the above equation becomes,

\begin{equation}\label{tempequn}
\frac{dT}{dt}=\frac{2}{3kn_{t}}(\Gamma^{heat}-\Lambda^{cool}-\frac{3}{2}kT\frac{dn_{t}}{dt})
\end{equation}
The last term on the right-hand side of this equation is the rate of change in the internal energy density associated with the rate of change of the total number of free particles in the gas. This is negligible when the gas is highly ionized; otherwise, it explicitly couples the level populations with the temperature equation. Note that in the absence of explicit time dependence, $u$ is constant, and $T$ is determined from $\Gamma^{heat}=\Lambda^{cool}$.







At low or intermediate ionization parameters, heating mainly comes from photoionization. The photoionization heating for an ion $i$ of an element $X$ is given by,

\begin{equation}
    \Gamma^{pho}_{i,X}=\int_{\varepsilon_{th}(i,X)}^\infty \sigma_{(i,X)}(\varepsilon) J_\varepsilon(R,t)n_{i,X}(R,t)(\varepsilon -\varepsilon_{th (i,X)}) \frac{d\varepsilon}{\varepsilon}
\end{equation}
where $\sigma_{(i,X)}(\varepsilon)$ is the photoionization cross-section (cm$^{2}$) and $\varepsilon_{th (i,X)}$ is the ionization threshold energy of i$^{th}$ ionic stage of element $X$. $J_\varepsilon(R,t)$ is the mean intensity of the ionizing radiation field, which has the unit of erg s$^{-1}$ cm$^{-2}$ erg$^{-1}$, $n_{i,X}(R,t)$ are the number density of atom or ion at $i^{th}$ energy level of an element $X$ to be ionized in the unit of cm$^{-3}$ and $R$ is the distance in the cloud from the source of ionizing photons.

The above equation can also be written as:
\begin{equation}
    \Gamma^{pho}_{i,X}=n_{i,X}(R,t) \gamma_{i,X}(R,t){\langle {\varepsilon_{photo(i,X)}} \rangle}
\end{equation}
with
\begin{equation}
    {\langle {\varepsilon_{photo (i,X)}} \rangle}=\frac{\int_{\varepsilon_{th(i,X)}}^\infty J_\varepsilon(R,t) \sigma_{(i,X)}(\varepsilon) (\varepsilon-\varepsilon_{th (i,X)}) \frac{d\varepsilon}{\varepsilon}}{\int_{\varepsilon_{th(i,X)}}^\infty J_\varepsilon(R,t) \sigma_{(i,X)}(\varepsilon) \frac{d\varepsilon}{\varepsilon}}
\end{equation}
and
\begin{equation}\label{eqngama}
    \gamma_{i,X}(R,t)=\int_ {\varepsilon_{th(i,X)}}^\infty  \sigma_{(i,X)}(\varepsilon) J_\varepsilon(R,t) \frac{d\varepsilon}{\varepsilon}
\end{equation}
where $\gamma_{i,X}(R,t)$ is the photoionization rate and ${\langle {\varepsilon_{photo(i,X)}} \rangle}$ is the mean kinetic energy of photo-ionized electron weighted by the photoionization cross-section for an $i^{th}$ ionic stage of an element $X$. The total heating rates contributed by all the ions and elements are now given by,

\begin{equation}
    \Gamma^{pho}=\sum_{X}\sum_{i}\Gamma^{pho}_{i,X}
\end{equation}

The mean intensity of radiation field $J_\varepsilon(R,t)$ for spherical geometry is given by,
\begin{equation}
   J_\varepsilon(R,t)=\frac{L_\varepsilon(R,t)}{4\pi R^2}
\end{equation}
where $L_\varepsilon(R,t)$ is the specific luminosity. Cooling comes from recombination, collisional ionization, and collisional excitation. We also include Compton heating, Compton cooling, and bremsstrahlung cooling in our calculation.

\subsection{Time Dependent Radiative Transfer}
The equation describing the time evolution of the radiation field in spherical symmetry is \citep{hat76,gar13},
\begin{equation}\label{rad}
    \frac{1}{c}\frac{\partial{L_\varepsilon(R,t)}}{\partial t}+ \frac{\partial{L_\varepsilon(R,t)}}{\partial R} = 4 \pi R^2 j_\varepsilon-\kappa_\varepsilon(R,t) L_\varepsilon(R,t) \end{equation}
where $R$ is the distance in the cloud from the ionizing source, $L_\varepsilon(R,t)$ is the specific luminosity of the ionizing source in erg s$^{-1}$  erg$^{-1}$, $j_\varepsilon$ is the local emissivity in erg s$^{-1}$ cm$^{-3}$ erg$^{-1}$, and $\kappa_\varepsilon(R,t)$ is the total extinction coefficient in cm$^{-1}$. In our model calculation, extinction comes from absorption only. Also, we do not include emission, and we assume all rays are radial, so the equation becomes:

\begin{equation}\label{transfer}
    \frac{1}{c}\frac{\partial{L_\varepsilon(R,t)}}{\partial t}+ \frac{\partial{L_\varepsilon(R,t)}}{\partial R} = -\kappa_\varepsilon(R,t) L_\varepsilon(R,t)
\end{equation}

\subsection{Timescales}
General solutions to equations (\ref{levpop}), (\ref{chargebalance}), (\ref{tempequn}) and (\ref{transfer}) require numerical calculation, and we will present these in subsequent sections. {\bf }However, simple analytic estimates help understand the general properties of time dependent photoionization.  
Following \cite{gar13}, we define the characteristic timescale affecting photoionization as:

\begin{equation}\label{eqn12}
    t_{pi}=\frac{1}{\gamma}
\end{equation}

\noindent where $\gamma$ is the photoionization rate as defined in equation (\ref{eqngama}).  
If the radiation field suddenly decreases, then the important characteristic time is the recombination time.

\begin{equation}\label{eqn13}
    t_{rec}=\frac{1}{n_e\alpha_r}
\end{equation}

\noindent where $\alpha_r$ is the recombination rate coefficient in units of cm$^3$ s$^{-1}$ and $n_e$ is electron number density.
Similarly, the collisional ionization timescale can be defined as: 

\begin{equation}\label{eqn14}
    t_{col}=\frac{1}{n_e\alpha_c}
\end{equation}
where $\alpha_c$ is the collisional ionization rate coefficient in units of ${\rm cm}^3\ {\rm s}^{-1}$. These definitions differ from those of \cite{gar13} in that equations \ref{eqn12} -- \ref{eqn14} are $per \ ion$ rather than per hydrogen atom. These timescales are intrinsic to each ion in the gas for a given radiation flux, temperature, and electron density.

Using these timescales, we can define the  ionization equilibrium timescale as the time it takes for the gas to go from one equilibrium state to another and can be estimated using the ionization balance equation (\ref{levpop}) for the simple case of the two-level atom, one bound level plus continuum with constant electron number density, the solution to this equation has a simple exponential character.  The characteristic timescale governing the approach to photoionization equilibrium is:

\begin{eqnarray}
    t_{ion}&=& \frac{t_{rec}t_{pi}}{t_{rec}+t_{pi}}
\end{eqnarray}
where $t_{ion}$ is characteristics ionization timescale. This shows that both the characteristics recombination and photoionization timescales affect the ionization equilibrium timescale.

The temperature equilibration timescale can be found using the energy balance equation (\ref{tempequn}). This can be rearranged to give the thermal equilibrium time as: 

\begin{equation}\label{thermtime}
    t_{temp}=\frac{\frac{3}{2}n_t k (T-T^E)}{\Gamma^{heat}-\Lambda^{cool}} 
\end{equation}
where $T$ is the equilibrium temperature corresponding to the initial flux and $T^E$ equilibrium temperature for the final flux. This shows that the timescale for the approach to thermal equilibrium depends on the heating and cooling rates.  These, in turn, depend on ionization and recombination rates and also on collisional rates affecting the excitation of atoms and ions.

The response of a cloud to a changing ionizing source is also affected by the rate at which photons are supplied. That is since each ionization event consumes a photon, the size of the region subject to photoionization cannot grow faster than the supply of photons.  Following \cite{gar13}, we call this the propagation timescale.  It can be defined by  equating the total number of ions in the volume of ionized gas, i.e., $4\pi R^3 n_{H}/3$, to the total number of ionizing photons emitted by the ionizing source in a time $t_{prop}$, $L_{ion} t_{prop}/\langle{\varepsilon_{prop}}\rangle $, where $\langle{\varepsilon_{prop}}\rangle$ is the mean ionizing photon energy  and $L_{ion}$ is the total ionizing luminosity.  Then it is easy to show that: 

\begin{equation} \label{eqn20}
    t_{prop} =  \frac{N_{H} \langle{\varepsilon_{prop}}\rangle }{3F_{ion}}
\end{equation}
where  $N_{H}$ is the total hydrogen column density and $F_{ion}$ is the total ionizing flux which is equal to $L_{ion}/(4 \pi R^2)$. Equation \ref{eqn20} is based on the assumption of instantaneous ionization.

The light travel time is given by,
\begin{equation}\label{eqn21}
    t_{light}=\frac{r}{c}
\end{equation}
where $r$ is the distance of a point in the cloud from the illuminating face and $c$ is the speed of light. This represents the obvious constraint that the signal representing the change in the ionizing radiation flux cannot travel faster than the speed of light.  It also illustrates a more general point:  that the response of a cloud to a changing light source will evolve on the {\it slowest} relevant timescale.

\subsection{Comparison of Timescales}
The ion fractions and other properties of a photoionized cloud will evolve with time subject to the timescales described in equations \ref{eqn12} -- \ref{eqn21}.  The net effect of the various timescales is that this evolution will happen at the slowest of the relevant timescales.  The timescales depend on the ionizing flux of radiation, the physical size of the cloud, and the gas density. In this section, we examine how  the various timescales vary with these quantities. In order to do so, we adopt a simplification:  that the radiation flux incident on the cloud is determined by simple geometric dilution of the radiation from a point source and that the size of the cloud is equal to the distance from the radiation source to the illuminated face of the cloud.  That is, we assume that the radiation flux incident on the cloud is $F_{ion}=L_{ion}/(4\pi R^2)$, where $L_{ion}$ is the intrinsic ionizing luminosity of the source, and that the cloud size is equal to $R$, the distance from the radiation source to the cloud illuminated face.  That is that the cloud column density is $N_H=n_HR$. {\it We emphasize that  these assumptions are adopted only within this section and only for illustration}. The numerical models in the subsequent section describe clouds with sizes and incident fluxes that do not obey these assumptions. Throughout the rest of the paper, the distance of a point in the cloud from the radiation source is denoted by R, and the distance of a point in the cloud from the illuminated face is denoted by r.

With these assumptions, we can express the various timescales entirely in terms of $R$, the cloud size and distance from the source, the cloud density $n_H$, and the source luminosity $L_{ion}$, which we take to be $L_{ion}=10^{44}$ erg s$^{-1}$ for this illustration.  We make estimates for atomic rates, which are based on a hydrogenic approximation, here assuming a nuclear charge $Z=8$ (note that in our subsequent numerical models, we use the accurate rates from the {\sc xstar} database).  We also further emphasize that the estimates shown here do not take into account attenuation of the radiation and so are most accurate at the illuminated face of the cloud. However, we do take account of geometrical dilution in our estimates. Deep within a cloud, most timescales will be longer, owing to diminished total X-ray fluxes.  Thus, these estimates can be considered to be somewhat optimistic, but they serve to illustrate the regimes where various effects are likely to be important.

\begin{figure*}[ht]
\centering
\includegraphics[width=0.75\textwidth]{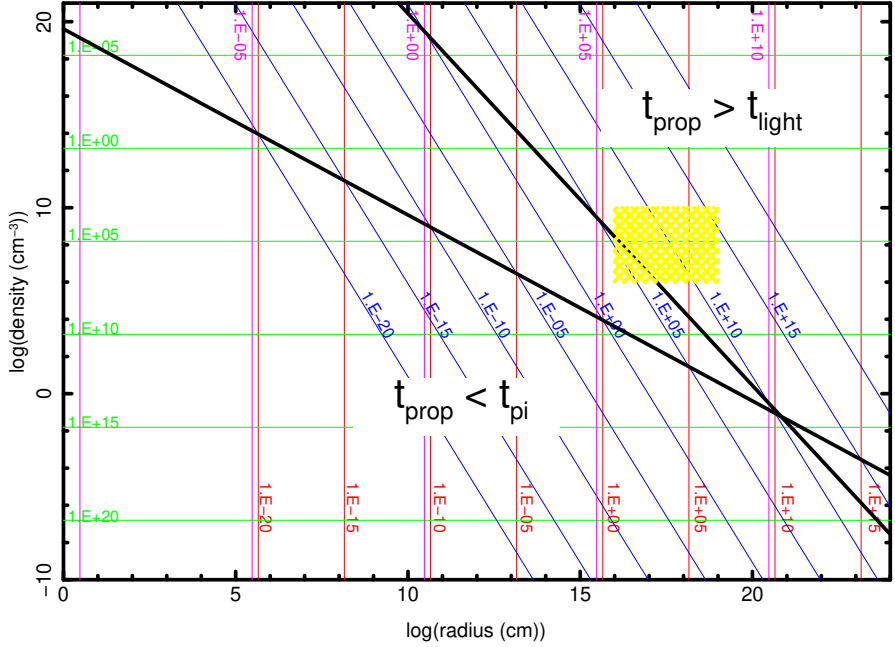}
\caption{Schematic illustration of the timescales relevant to a photoionized plasma in the radius and gas density (R,$n_H$) plane.  Estimates for rates are based on a hydrogenic approximation.  Colors label curves of various timescales: blue = propagation time; red = photoionization time; purple = light travel time; green = recombination time. Assumes a nuclear charge $Z = 8$ and a central source luminosity $L_{ion} = 10^{44}$ erg s$^{-1}$, crudely corresponding to a warm absorber.  The yellow hatched region shows the approximate range of $R$ and $n$ expected for typical warm absorbers.}
\label{fig01}
\end{figure*}

Figure \ref{fig01} illustrates the comparison between timescales defined in the previous subsection as a function of radius and density. Colors label curves of constant values for various timescales: blue = propagation time; red = photoionization time; purple = light travel time; green = recombination time.  Also shown are lines marking the boundaries of the regions where various timescales dominate.  In each region of the diagram, we expect the behavior of the gas to be determined by the slowest timescale.  In the lower-left of the diagram,  the region  marked $t_{prop} < t_{pi}$, the photoionization timescale is longer than the propagation time.  In this region, the rate of supply of photons from the central source exceeds the rate at which photoionization proceeds.  If so, neglect of time dependent transfer is justified. In the upper right of the diagram, in the region marked $t_{prop} >  t_{light}$, the propagation time is longer than the light travel time.  In this region time dependent transfer must be taken into account. We emphasize that in regions where $t_{prop} < t_{light}$ the movement of the ionization front (Hereafter ionization front is abbreviated as IF) does not exceed the speed of light; rather, the speed at which the IF moves is {\it determined} by the speed of light, and the rate of supply of photons does not limit the speed. In reality, the boundaries between various regions  are not sharp, as indicated in the diagram, because the transition from the dominance of one timescale or another is not sudden.  But they serve as useful indicators of where various physical processes dominate the time evolution of photoionization.  In our calculations in subsequent sections, of course, we include all relevant processes.

Figure \ref{fig01} also shows that: (i) Timescales $\geq 10^3$ sec are characteristic in the region ${\rm log}(R)\geq$17.  Such timescales are typical for AGN variability, and such distances are plausible locations for warm absorber clouds.  This illustrates  that the time needed for the cloud to respond to the central source variability is not short compared to the source variability timescale. (ii) The propagation time exceeds the photoionization timescale eg. for densities ${\rm log}(n_H)\geq$1 at ${\rm log}(R)$ = 18.  Since this includes a wide range of likely warm absorber cloud densities, this illustrates that time dependent radiative transfer is important for warm absorber clouds. (iii) The light travel time exceeds the propagation timescale eg. for densities ${\rm log}(n_H)\geq$ 10 at ${\rm log}(R)$ = 18. Such conditions are somewhat greater than typically assumed for warm absorber clouds but are not generally ruled out.  Thus light travel time effects must also be included under such conditions.  (iv) At still greater densities (${\rm log}(n_H)\geq 10$), the photoionization time exceeds the recombination time, and the gas will be approximately neutral.

This demonstrates that, for the likely conditions in a warm absorber, characteristic timescales affecting the response of the gas to variability in the continuum are $\sim 10^3$ sec and greater.  And that finite photoionization, recombination, propagation, and light travel time effects must all be taken into account.

\section{Computation and Modeling} \label{s3}

In this section, we describe the computational details of our models:  the description of the ionization and heating of the gas, the discretization of the equations, the numerical solution method, and the values of the physical quantities used in the calculations.  

\subsection{Heating and Cooling Rates}\label{subsec3.1}
As discussed in section \ref{sec2}, the temperature of the gas depends on heating and cooling due to various atomic processes. For equilibrium models, these processes have been explored by many authors, e.g., \cite{kro81, kal04, roz14}. This shows that the heating and cooling rates have a characteristic behavior; when the ionization parameter is large, the gas is fully ionized, and the heating and cooling rates are dominated by Compton scattering and bremsstrahlung, resulting in an equilibrium temperature that depends primarily on the shape of the illuminating spectrum, and is $\sim 10^7$ -- $10^8$ K. When the ionization parameter is small the gas becomes nearly-neutral. The rates are dominated by hydrogen and helium atomic processes. In this limit, the temperature is $\sim 10^4$ K. At intermediate ionization parameters, the rates are dominated by contributions from many trace ions and elements, and the equilibrium temperature takes intermediate values. This behavior is described by the `s curve,' which is the locus of the equilibrium temperature in the ionization parameter-temperature plane. This behavior has been extensively modeled in the literature \citep{bott2000, kro81,cha12, mat87}.

Time dependent calculations require the simultaneous solution of the equations describing the ionic level populations, temperature, and radiation field for all level populations,  spatial zones, and photon energies.  This corresponds to a very large number of simultaneous ordinary differential equations; the standard {\sc xstar} database has $\sim 3 \times 10^4$ energy levels.  This necessitates a simplification of this system of equations to make it tractable.  That is, we must simplify the energy level structure of the ions in the gas. To do this, we have created an atomic database that describes each ion using  three levels; the ground level, one excited bound level, and the continuum (ionized) level.  We include all of the ions that are in {\sc xstar} in the newly created database, though the simulations in this paper include only a subset of them.  

Using our simplified energy level scheme, our  atomic database contains all the important atomic processes affecting these levels.  For the one bound excited level, we adopt an {\it ad hoc} description:  the level energy is 0.8$\times E_{th}$ where $E_{th}$ is the ionization potential of the ion. We include electron impact collisional excitation to the level for the ions H$^0$, He$^0$, and He$^+$ with an effective collision strength (upsilon) with a value of 1.4 to 2.7 depending upon temperature.

This ad hoc energy level structure is created primarily to obtain thermal balance. However, we also use it for calculating the spectrum. We solve the time-dependent coupled differential equations of ionization balance, heating and cooling, and radiative transfer simultaneously and self-consistently. The absorption spectra we are calculating mostly depend upon the ground state populations, so the excited state structure and level population are relatively unimportant.

Since the ionization and excitation states of the gas are coupled with the time-dependent radiation field, the ionization fractions, and temperature and radiation field quantities are determined uniquely for each specific problem and associated input quantities.  For this reason, existing published or tabulated cooling data cannot be applied to our calculations.  A full self-consistent calculation is needed, and our simplified energy level structure is required in order to make this tractable. 

\begin{figure*}[ht]
\centering
\includegraphics[width=0.7\textwidth]{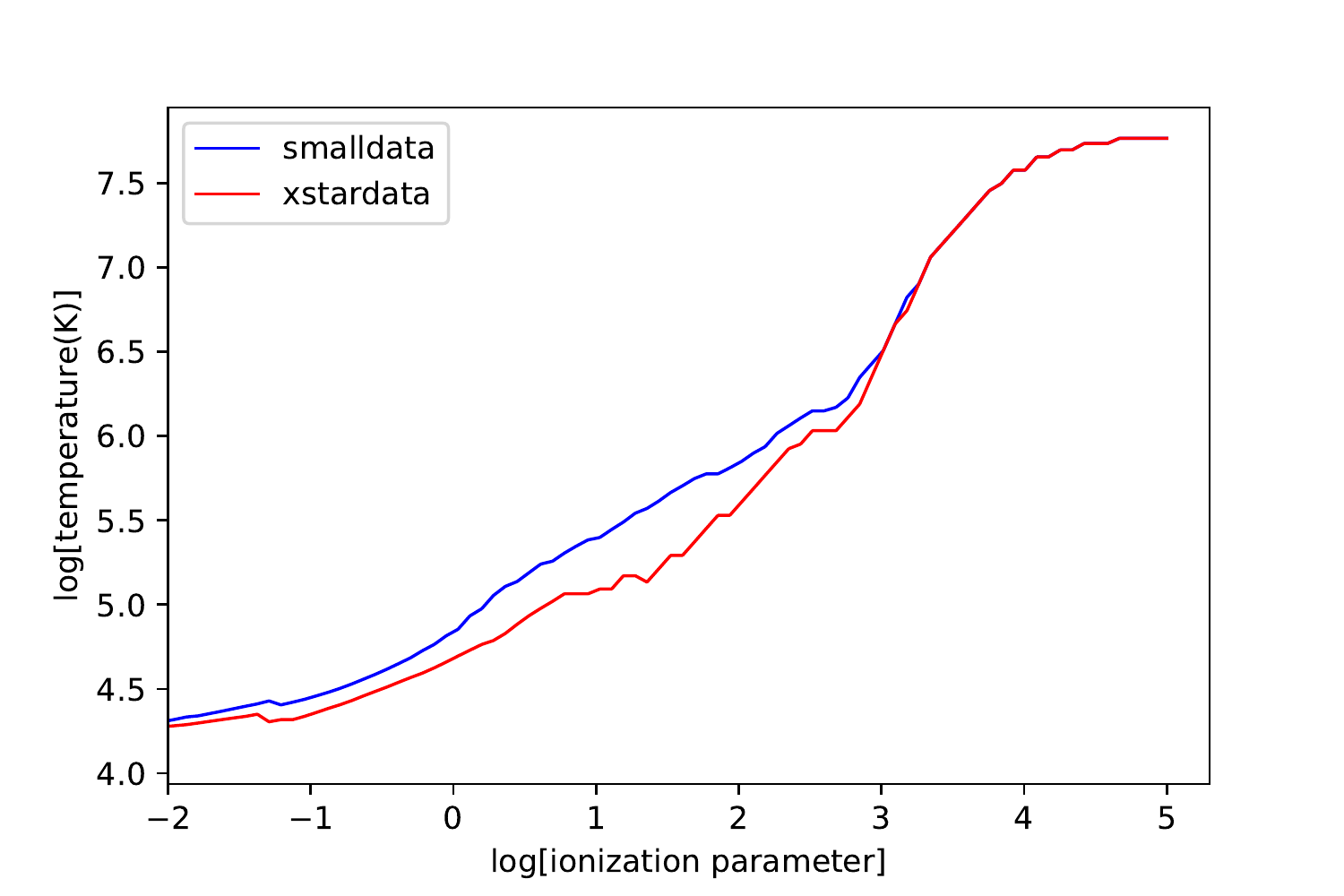}
\caption{Comparison of equilibrium temperature obtained using two different databases. The red curve corresponds to the standard {\sc xstar} database, and the blue curve is for the small database we created for this project.}
\label{fig2}
\end{figure*}

In figure \ref{fig2}, we  show the comparison of the equilibrium temperature vs. ionization parameter (the `s curve') calculated using the small database used in our time dependent calculation compared with the standard {\sc xstar} database. The red curve is obtained from {\sc xstar} and its database \citep{kall01} and the blue curve is obtained from running {\sc xstar} using our small database. This shows good agreement at high and low ionization parameters and disagreement of up to $\sim$ 0.5 dexes at intermediate values of ionization parameter. Since parts of our calculations reside in the intermediate ionization parameter range, some error is introduced by the use of the small database. However, this difference is smaller than other  prevalent uncertainties in the equilibrium temperature.  For example, comparable changes in the s curve are associated with varying assumptions about the shape of the illuminating spectrum and about the chemical composition of the gas.  Both of these quantities are poorly constrained observationally \citep{kro81,cha12,mat87}.


\begin{figure*}[ht]
\centering

\includegraphics[width=0.9\textwidth]{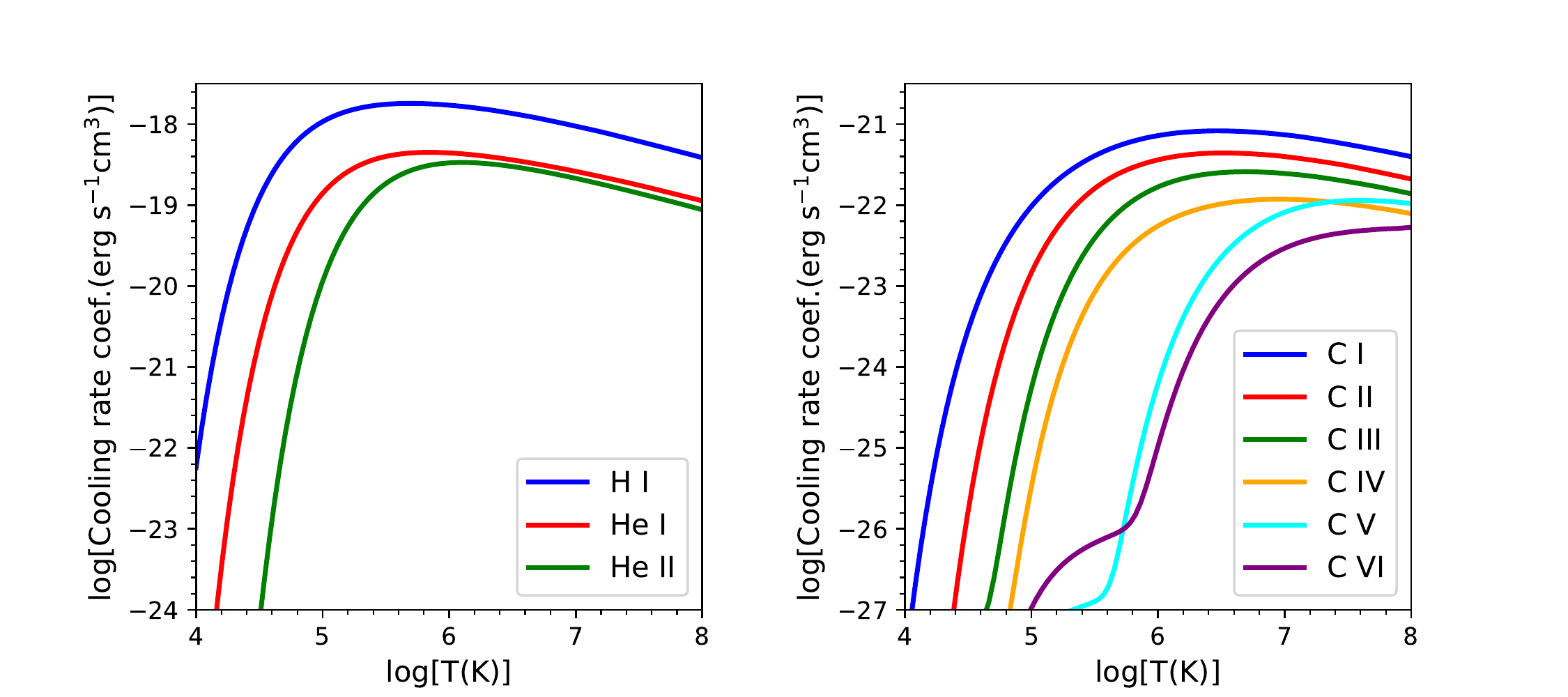}
\caption{The left panel shows the cooling rates per ion for H I, He I, and He II, and the right panel shows the cooling rates per ion for carbon ions. }
\label{hc}
\end{figure*}

Since we have developed a brand new non-equilibrium code, we  compare some results against the results obtained by using more mature codes. The comparison shown in figure \ref{fig2} represents the most appropriate test of our small database since the full {\sc xstar} database used in this comparison has been extensively tested against other similar codes \citep{med2016}.  Another test is shown in 
Figure \ref{hc}, which shows the cooling rates for different ions of hydrogen, helium, and carbon vs. temperature for a coronal gas.  The cooling rates are calculated by neglecting photoionization and keeping the ion fraction values unity. These results can be compared with the results obtained by \cite{gnat2012}; 
This demonstrates that even though our study focuses on photoionized plasmas, the code can be applied to  other problems, for example where the equilibrium approximation is invalid due to other causes, such as heating by shock waves or sudden expansion or contraction of the gas.

\subsection{Finite Difference Method}
We solve equations (\ref{levpop}), (\ref{chargebalance}), (\ref{tempequn}), and (\ref{transfer}) by discretizing using the  finite difference method. We divide cloud thickness, time, and incident spectrum into bins and calculate the physical quantities of interest in each bin. The set of equations to be solved have a wide range of timescales. For the warm absorber case, the ion fractions change over timescales ranging from $10^3$ s to $10^5$ s. On the other hand, the temperature evolves over a longer timescale than the photoionization timescale. The transmitted spectra change over a timescale of $\sim 10^7$ s. Thus, combining all these equations leads us to a set of stiff differential equations. Since our equations are stiff, we implement the backward Euler differentiation method. In this method, we calculate the functional value at a one time-step ahead. This method leads us to have a set of non-linear equations and we use the modified Newton method to solve it \citep{bro89}.

Adopting the definition of implicit differentiation, we find the corresponding equation governing the time evolution of the level populations (\ref{levpop}):

\begin{equation}
\label{eqn5}
    n_{i,X}^{t+1}-\Delta t \left[\sum_{j=1}^p n_{j,X}^{t+1} R_{ji}^{t+1}-\sum_{k=1}^p n_{i,X}^{t+1} R_{ik}^{t+1}\right]=n_{i,X}^{t}
\end{equation}
where the symbols are the same as defined in equation (\ref{levpop}) above, except for the superscript $t$, which indicates the time step.

In the same way with approximation, $n_t=n_H+n_e$ where $n_e$ is the electron number density, equation (\ref{tempequn}) becomes; \footnote{superscript t is for time step, and $n_t$ is the total number density of particles in the ionized gas.}

\begin{equation}
\label{eqn6}
\begin{split}
T^{t+1}- \frac{2\Delta t} {3kn_{t}}\left[\Lambda^{(heat)t+1}-\Gamma^{(cool)t+1})\right]\\
+T^{t+1}\frac{(n_{t}^{t+1}-n_{t}^{t})}{n_{t}}=T^t
\end{split}
\end{equation}
The radiative transfer equation (\ref{transfer}) becomes,



\begin{equation}
\begin{split}
\label{eqn7}
    L_\varepsilon^{t+1}(R,t)+ c\Delta t \kappa_\varepsilon^{t+1} (R,t) L_\varepsilon^{t+1}(R,t)+\\c\Delta t \frac{(L_\varepsilon^{t+1}(R+\Delta R,t)-L_\varepsilon^{t+1}(R,t))}{\Delta R}=L_\varepsilon^{t}(R,t)
\end{split}
\end{equation}

The initial values of the right-hand side of the equations (\ref{eqn5}), (\ref{eqn6}), (\ref{eqn7}) are obtained from the initial equilibrium calculation obtained with {\sc xstar}. With this, we end up with a set of nonlinear equations in variables of level population, temperature, and specific luminosity at different energy bins. The main task is to solve this system of coupled nonlinear equations simultaneously and self-consistently. The details of how it is solved are discussed in appendix \ref{if0}.  The input parameters and assumptions are described in the subsection below. 

\subsection{Model Parameters}
In this paper, we present models of two basic types. First is a low-density gas model designed to resemble an H II region. Second, and explored in more detail, are models for warm absorbers.  For both model types, we investigate the properties of gas when it is exposed to a sudden change in the ionizing incident flux, either a step up or a step down. This is idealized compared with some observations, but it serves to illustrate the general behavior, and such models can be used to synthesize the behavior under more realistic assumptions about the variation of the illuminating flux. Since we use logarithmic time steps., it is convenient to start with a small but non-zero time.  We take this time to be 10 s. The step up or step down is at a time of 20 s for the warm absorber models.  For the H II region model, the step up occurs at 50 s.

The H II region model is the same as the model presented by  \cite{gar13}, but with improvements:  the inclusion of more elements and the inclusion of geometrical dilution appropriate to an extended cloud.  We present this model in order to compare with \cite{gar13} and to illustrate the effects of including more chemical elements and the geometric dilution factor.
\begin{figure}[h!]
\centering
\includegraphics[width=.4\textwidth]{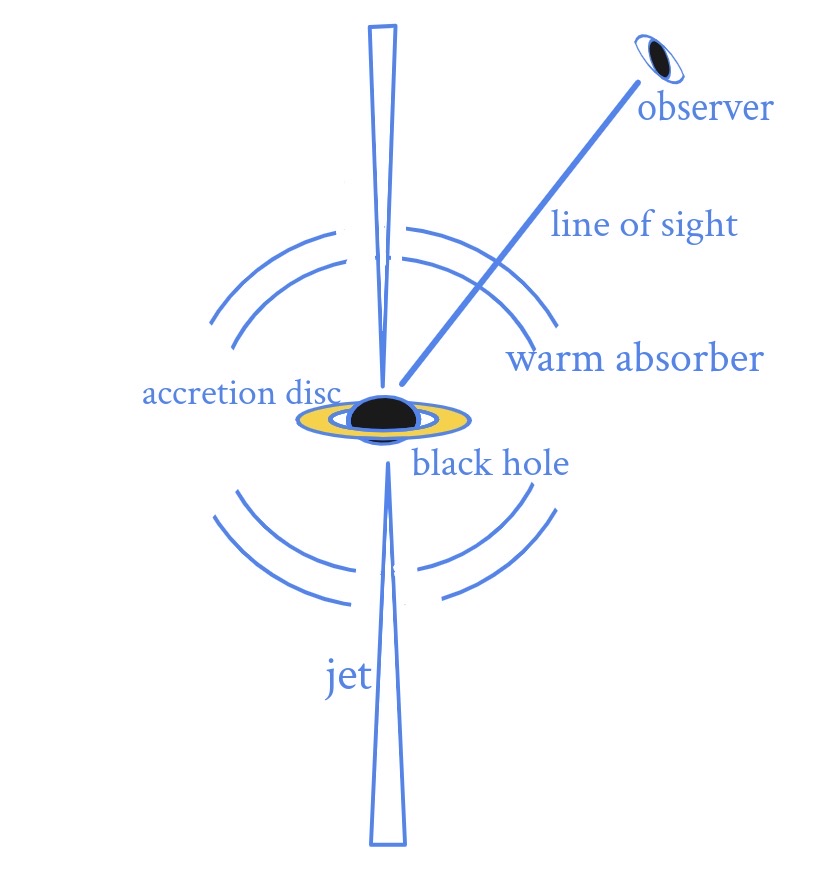}
\caption{Cartoon (not in scale) showing different components of the central part of an AGN.}
\label{fig1}
\end{figure}

In the H II region model, we adopt a density = $10^4$ cm$^{-3}$,  luminosity of the illuminating source = $10^{32}$ erg s$^{-1}$, ionization parameter $\xi \approx$ 1 erg cm s$^{-1}$, initial ionizing flux ($F_{ion,1}$) = 795 erg s$^{-1}$ cm$^{-2}$, final ionizing flux ($F_{ion,2}$) = 2390 erg s$^{-1}$ cm$^{-2}$. The inner radius is $10^{14}$ cm, the outer radius of the cloud is $10^{18}$ cm, and column density of $10^{22}$ cm$^{-2}$ and we include elements: H, He, C, O, Si and Fe. The cloud is divided into 80 spatial zones, spaced  logarithmically. The initial and final times are 10 s and $10^{10}$ s, and the total time is divided into 40 logarithmic time grid points (though the ODE solver subdivides within these intervals as needed). The spectral energy distribution of the central source is  a power law spectrum with an energy index of -1. The spectrum is divided into 50  grid points spaced logarithmically between 1 eV and 10 keV. If all the neutral and ionic stages of included elements are considered, we have 57 ions. Each of the ions is considered to have three energy levels. However, the ionized level of the atom at i$^{th}$ ionization stage corresponds to the ground state of (i+1)$^{th}$ ionization stage of the same ion. This results in each element having $2Z+1$ energy levels, where $Z$ is an atomic number. Taking this into account, we have 120 level population equations for the elements included. In addition, we have one equation for temperature, one for electron fraction, and 50 radiative transfer equations, one for each energy,  giving 172 ordinary differential equations for each spatial zone. Since we have 80 spatial zones, we will have a total of 13,760 differential equations. This is then solved for time $t+1$ using differential equation solver DVODE \citep{bro89}. This process is repeated until we reach the final time points. The final time for this model is $10^{10}$ s.

\begin{table*}[t!]
    \centering
    \begin{tabular}{|c|c|c|c|c|c|c|c|}
    \hline
    Initial Flux($F_{ion,1}$) & Final Flux($F_{ion,2}$) & Density($n_H$)  & Initial $(\xi)$ & \multicolumn{4}{|c|}{Column Density($N_H$)}\\
    
     (erg  cm$^{-2}$ s$^{-1}$) & (erg  cm$^{-2}$ s$^{-1}$) & (cm$^{-3}$) & (erg cm s$^{-1}$) &\multicolumn{4}{|c|}{(cm$^{-2}$)}\\\hline
     & & & &0.9$\times$10$^{23}$&1.3$\times$10$^{23}$&1.4$\times$10$^{23}$&2.0$\times$10$^{23}$\\\hline
     & & & &\multicolumn{4}{|c|}{Model}\\\hline
     3.2$\times$10$^7$&9.6$\times$10$^7$&10$^7$&40&\textbf{13}&\textbf{33}&&\\\hline
     4.0$\times$10$^7$&1.2$\times$10$^8$&10$^7$&50&\textbf{73}&\textbf{93}&&\\\hline
     6.5$\times$10$^7$&1.9$\times$10$^8$&10$^7$&81&&&\textbf{14}&\textbf{34}\\\hline
     8.5$\times$10$^7$&2.5$\times$10$^8$&10$^7$&100&&&\textbf{74}&\textbf{94}\\\hline
     6.5$\times$10$^{11}$&1.9$\times$10$^{12}$&10$^{11}$&81&&&\textbf{15}&\textbf{35}\\\hline
     8.5$\times$10$^{11}$&2.5$\times$10$^{12}$&10$^{11}$&100&&&\textbf{75}&\textbf{95}\\\hline
     9.6$\times$10$^7$&3.2$\times$10$^7$&10$^7$&120&\textbf{16}&\textbf{36}&&\\\hline
     1.2$\times$10$^8$&4.0$\times$10$^7$&10$^7$&150&\textbf{76}&\textbf{96}&&\\\hline
    \end{tabular}
    \caption{Table showing  warm absorber model parameters. Boldface numbers represent the model names. In our discussion, we focus on the detailed behavior of models 73, 74, 75, and 76. We identify them for convenience with the names `baseline model' for model 73 (baseline model), `high flux high column model' for 74, `high density model' for model 75, and `step down flux model' for model 76.}
    \label{tab1}
\end{table*}

 The second type of model we consider is a warm absorber model. Figure \ref{fig1} is a cartoon schematically showing the location of the warm absorber and illuminating source. Warm absorbers intercept the radiation coming from the central part of the AGN. This radiation is absorbed as it travels through the gas. Viewing the central ionizing source through the line of sight shown in the figure, we observe the absorption spectrum \citep{blu05}.  Simple estimates show that the thickness of the gas is small compared to its distance from the source. We consider two different gas densities $10^7$ cm$^{-3}$ and $10^{11}$ cm$^{-3}$ for different models, initial ionization parameter $\sim$ 100 erg cm s$^{-1}$, luminosity of the source $10^{44}$ erg s$^{-1}$ and column density $\sim 10^{23}$ cm$^{-2}$ and include elements: H, He, C, O, Si and Fe. The cloud is divided into 40 spatial zones, spaced  logarithmically. The initial and final times are 10 s and $10^{9}$ s, and the total time is divided into 40 logarithmic time grid points except model 76. These grid points are separated by $\Delta{\rm log}t=0.23$. In model 76, we have used two types of time steps. Logarithmic time step is being used up to $10^7$ s with 20 grid points and linear time step of $\approx 1.3\times10^7$ s after that with 79 time grid points. The spectral energy distribution of the central source is  a power-law spectrum with an energy index of -1. The spectrum is divided into 100 grid points spaced logarithmically. This leads to us having 8,880 differential equations to be solved, including electron and temperature equations for each time point. The model parameters for different WA models are listed in tables \ref{tab1} and \ref{dist_tab}.  Table \ref{dist_tab} shows that the geometrical thickness of the warm absorber models is small compared with the distance from the radiation source.  The models $15$, $35$, $75$, and $95$ are warm absorber models where the final flux is lowered by a factor of 3, and we call these models step down  models. All other models assume a  step up in flux by a factor of 3 and are called step up flux models.

In the rest of this paper, we present the results of all of our models, but we focus on the detailed behavior of models 73, 74, 75, and 76. We identify them for convenience with the names `baseline model' for model 73 (baseline model), `high flux high column model' for 74, `high density model' for model 75, and `step down flux model' for model 76.

We tested the warm absorber model taking 40 and 80 spatial zones. The results remain nearly the same for both choices. This is primarily due to the implicit calculation we include in solving time-dependent equations. Hence we chose 40 spatial zones to make the computation tractable.

Our models employ an energy grid  using  100 energy grid points spaced logarithmically over the energy range  1-$10^4$~eV  ({\sc xstar} also includes a very small number of energy bins between $10^4$ and $10^6$ eV in order to describe the Compton heating accurately). As a result, the energies of spectral features, lines, and edges, are binned into rather crude bins, approximately 100 eV wide at 1 keV, and this limits the precision of the energy scale for the results shown in this paper.

\subsection{Assumptions}
This work is the first step toward exploring photoionized plasmas in non-equilibrium conditions. In order to make our results general and to illustrate the general behavior of time dependence, we have made several key approximations, which are as follows:
\begin{enumerate}
    \item We approximate the gas to be static, with no expansion of the gas and no velocity gradient. The shift in the spectral features due to these effects will be small compared with our spectral resolution. On the other hand, for a supersonic flow, such as in a warm absorber, the bulk forces, whatever they are, dominate over internal dynamics in the cloud \citep{proga22}. We acknowledge this effect can be important. However, we have chosen to make simple general models in which the gas motion is neglected. We can calculate models which can be refined for specific situations.
    \item We assume the outflowing gas has constant density throughout. This assumption is partially justified by the physical thickness of the cloud, which is less than $10\%$ with respect to the distance from the ionizing source, as shown in table \ref{dist_tab}. If so, the imprint of any  large-scale spherical flow will be small. However, we acknowledge that  there may be a density gradient in the case of warm absorbers associated with dynamical effects such as ram pressure, which will be considered in future work.
    \item We have not included emission in our radiative transfer calculation. This approximation is justified if the covering fraction of the absorber relative to the central source is small; this appears to be the case for most observed warm absorbers. For most observed warm absorber spectra, emission features are weak or absent  \citep{kas01,kas02}. However, including emissions is on our list of tasks for future work since the emission is likely to be important in other time dependent applications.
\end{enumerate}

\begin{table}[t]
    \centering
    \begin{tabular}{|c|c|c|c|}
    \hline
    Model name & $R_0$ & $\Delta R$ & $\Delta R / R_0$(in $\%)$\\\hline
    \textbf{13} & 5.0$\times$10$^{17}$&9.0$\times$10$^{15}$& 1.8\\\hline
    \textbf{33} & 5.0$\times$10$^{17}$&1.3$\times$10$^{16}$& 2.6\\\hline
    \textbf{73} & 4.5$\times$10$^{17}$&9.0$\times$10$^{15}$& 2.0\\\hline
    \textbf{93} & 4.5$\times$10$^{17}$&1.3$\times$10$^{16}$& 2.9\\\hline
    \textbf{14} & 3.5$\times$10$^{17}$&1.4$\times$10$^{16}$& 4.0\\\hline
    \textbf{34} & 3.5$\times$10$^{17}$&2.0$\times$10$^{16}$& 5.7\\\hline
    \textbf{74} & 3.0$\times$10$^{17}$&1.4$\times$10$^{16}$& 4.4\\\hline
    \textbf{94} & 3.0$\times$10$^{17}$&2.0$\times$10$^{16}$& 4.5\\\hline
    \textbf{15} & 3.5$\times$10$^{15}$&1.4$\times$10$^{12}$& 0.038\\\hline
    \textbf{35} & 3.5$\times$10$^{15}$&2.0$\times$10$^{12}$& 0.057\\\hline
    \textbf{75} & 3.0$\times$10$^{15}$&1.4$\times$10$^{12}$& 0.046\\\hline
    \textbf{95} & 3.0$\times$10$^{15}$&2.0$\times$10$^{12}$& 0.065\\\hline
    \textbf{16} & 2.9$\times$10$^{17}$&9.0$\times$10$^{15}$& 3.1\\\hline
    \textbf{36} & 2.9$\times$10$^{17}$&1.3$\times$10$^{16}$& 4.5\\\hline
    \textbf{76} & 2.6$\times$10$^{17}$&9.0$\times$10$^{15}$& 3.5\\\hline
    \textbf{96} & 2.6$\times$10$^{17}$&1.3$\times$10$^{16}$& 5.0\\\hline
    
    \end{tabular}
    \caption{Table showing the distance of the illuminated face of the  absorber from the ionizing source ($R_0$), the geometrical thickness of the absorber ($\Delta R$), and the ratio of the thickness of the absorber and distance of absorber from ionizing source in percent. This does not include the H II region model which is described in the paragraph in subsection 3.3}
    \label{dist_tab}
\end{table}

\section{Results} \label{sec4}
In this section, we present model results for  the H II region  and warm absorber models. We present  the time evolution of the ion fractions, temperature, and electron fraction in the gas. We also present the evolution of the emergent spectrum and discuss the effect of time dependent photoionization on the power density spectrum.

\subsection{H II Region Model}
\begin{figure*}[ht]
\centering
\includegraphics[width=1.\textwidth]{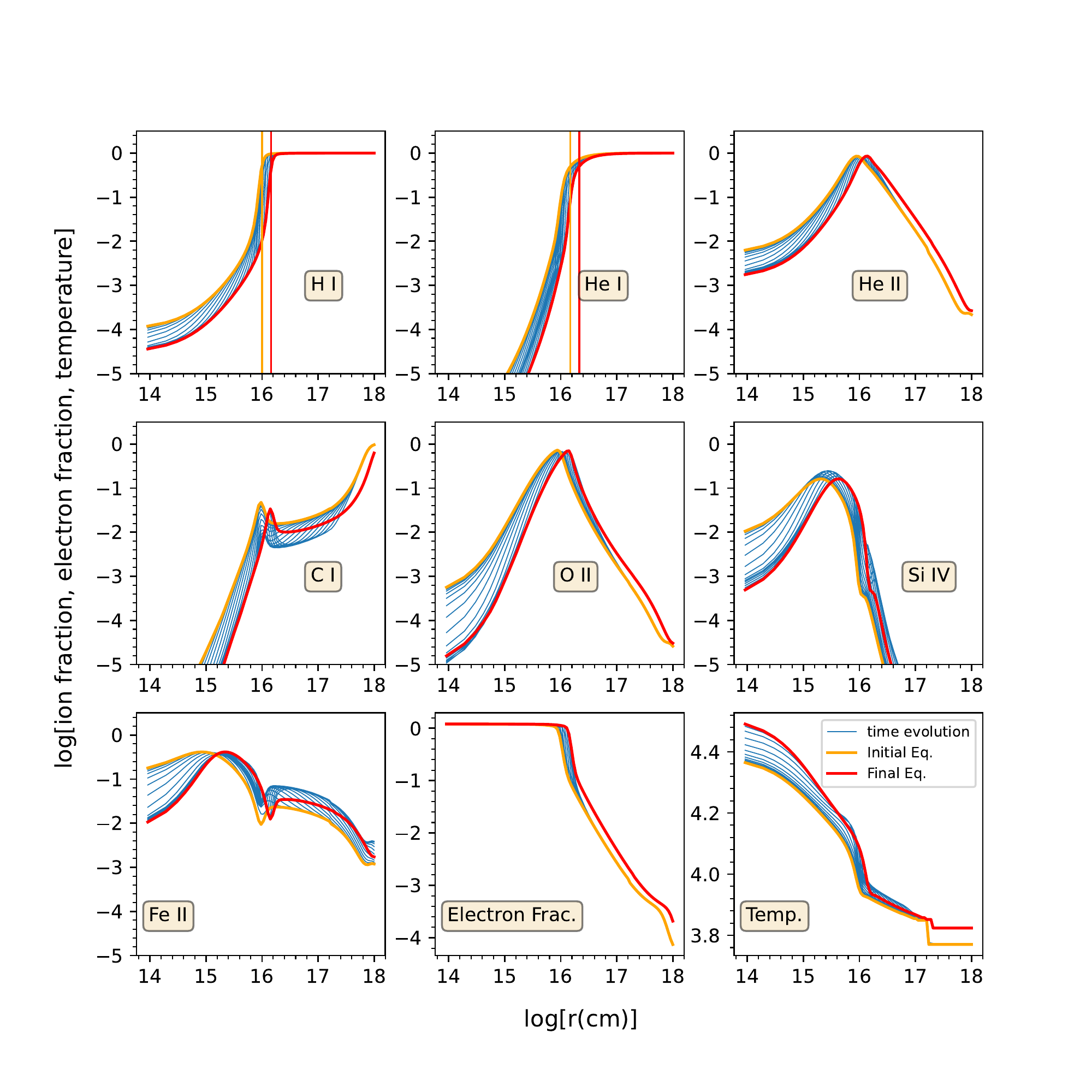}
\caption{Figure showing the evolution of ion fraction, electron fraction, and temperature profiles in H II region model of the cloud with density $n=10^4$ cm$^{-3}$. The orange and red curve represents (the lower flux 795 erg s$^{-1}$ cm$^{-2}$) initial and final equilibrium (high flux, which is 3 times the lower flux; 2390 erg s$^{-1}$ cm$^{-2}$) while the intermediate blue lines represent the time evolution between the two equilibria. The x-axis represents the depth r in the cloud from the illuminated face in cm, while the y-axis represents the ion fraction, electron fraction, and temperature. The initial and final IFs are marked with the orange and red color vertical lines for  H~I and He I ions.}
\label{h2fig}
\end{figure*}

We first present the results of the time dependent response in the case of the H II region model. We adopt the same conditions as used by  \cite{gar13} in order to facilitate comparisons with that paper. However, that paper included only hydrogen with ground and continuum levels, while we included the elements H, He, C, O, Si, and Fe with one additional excited level.

Figure \ref{h2fig} shows the evolution of ion fraction, electron fraction, and temperature profiles for this model. In this figure, we  include representative plots of ion fractions for some ions of hydrogen, helium, oxygen, silicon, and iron vs. depths in the cloud. The orange curve represents the initial equilibrium value. The rise in flux leads the gas to heat from photoionization, and the ionization of the gas is changed. The blue lines represent the intermediate state of the gas at various times during its evolution. The few initial times are: 10 s, 17 s, 29 s, 49 s, 83 s, 141 s, and a few times close to the final time are $2.0\times 10^9$ s, $3.4\times 10^9$ s, $5.8\times 10^9$ s, $10^{10}$ s. These lines are separated by $\Delta log(t) = 0.23$. Figure \ref{h2fig} shows that the changes in the ion fractions and temperature are not simultaneous at all depths but are delayed due to the fact that:  (i) the radiation field is more diluted at greater distances from the source, and so the photoionization time is longer and (ii) the propagation time is longer deeper in the cloud. Of these, the most important is the propagation time. When the gas near the face receives the increased flux, its  ionization rate increases, and it becomes more transparent. The gas deeper in the cloud receives more flux as time passes, but with a delay imposed by the photoionization time and propagation time. This process continues until the gas achieves a final equilibrium dictated by the changed flux. The final equilibrium ion fractions, electron fraction, and temperature are shown as the red curve in figure~\ref{h2fig}.
The temperature behavior is different on the IF. There is a sudden increase in the temperature surpassing the high flux equilibrium value. This increase in temperature happens because the low-energy photons are absorbed earlier in the ionized part of the gas, and hence mostly high-energy photons reach the IF. This is called the hardening of ionizing radiation. There, it ionizes the neutral atoms and releases more energetic electrons than it did in the ionized part of the cloud. This causes a rise in the temperature. This is seen in the temperature plot of figure \ref{h2fig} at $\sim 10^{16}$ cm. However, at a later time, the plasma becomes more ionized, and recombination cooling rates increase. This drives the temperature of the gas to equilibrium, gradually lowering the temperature. Hardening of ionizing radiation should also exist in the equilibrium calculation. Since the ionizing flux changes by the factor of 10,000 to a point near the IF with respect to the flux at the face, this effect is not so visible in the equilibrium calculation. This process, in principle, does not influence the gas outside the IF. However, the initial IF moves deeper into the cloud because of the increased flux.

The first few panels of the figure \ref{h2fig} show ion fractions vs. depths. The ion fractions slowly increase with depth and change abruptly at the IF. Hydrogen beyond the IF is predominantly neutral. The shape of the ion fraction distribution for neutral helium looks similar to hydrogen. He II, O II, and Si IV  evolve in different patterns. Ion fractions for these ions increase with depth in the cloud, reach a maximum at the IF, and then  decrease at depths beyond the IF. This  decrease is due to recombination into lower ionization stages. Fe II  has a  different shape: the ion fraction at first increases, becomes maximum and decreases again up to the IF. This may be due to the hardening of the radiation. The more energetic photons ionize the Fe II up to the IF, and the ion fraction increases  slightly beyond the IF due to the ionization of Fe I to Fe II by leftover high-energy photons. The electron fraction panel shows the electron number density distribution and clearly reveals the change in the location of the IF due to the flux increase. The last panel represents the temperature structure in the gas. The axes of the plot are in log scale. The temperature at the face of the cloud for this model is $\sim$ 23,000 K for the initial flux and $\sim$ 32,000 K for the final flux value.

For the H II region model, the time dependent solution does not converge to the steady state solution in the neutral region beyond $4 \times 10^{16}$ cm. This is because absorption has reduced the flux in this region so that the ionization timescale is long.  We choose to focus our H II region model results on the ionized region of the cloud, where the ionization and propagation timescales are shorter.  We have verified that running this H II region model for longer times ($10^{14}$ s or longer) results in convergence in the neutral cloud region. However, plots showing the neutral region time dependence necessarily skip over the time dependence in the ionized region, so we have chosen not to show this.


Our model differs from that of \cite{gar13} in a few ways. Since the gas extends from $10^{14}$ - $10^{18}$ cm from the ionizing source, geometrical dilution plays an important role in the ionization state of the cloud at different depths. Whereas \cite{gar13} did not include geometrical dilution in their calculation and, therefore, implicitly assumed the gas to be in a  plane-parallel configuration. The inclusion of dilution significantly decreases the flux far from the boundary of the cloud. This accounts for the  differences in the  curves for hydrogen bound level population and temperature between our work and \cite{gar13}. We have the initial IF at $\sim 10^{16}$ cm and no flat region near the boundary. In contrast, the  \cite{gar13} model has an IF at $\sim 4.0 \times 10^{16}$ cm and has a long flat region initially whereas we found the IF for a changed flux is at around $1.2 \times 10^{16}$ cm and \cite{gar13} have the changed IF at $\sim 1.3 \times 10^{17}$ cm. 
The shift in the IF between the initial and final equilibria is greater in \cite{gar13} for the same reasons. Another difference is the elements included. They included only hydrogen, whereas we included hydrogen, helium, carbon, oxygen, silicon, and iron in the simulation. The existence of a neutral region in our H II region model suggests that diffusely emitted radiation may be important in determining the ionization structure of the gas. We will explore this possibility in future work.

In appendix \ref{if1}, we present analytic estimates which  predict the IF for hydrogen at $1.0 \times 10^{16}$ cm and $\sim 1.5 \times 10^{16}$ cm. Our numerical results are close to the analytic results. The numerical results show the  IF for $He^0$ for initial flux is at $1.5 \times \sim 10^{16}$ cm and at $\sim 2.14 \times 10^{16}$ cm for changed flux. The analytic results  in appendix \ref{if1} predict initial IF is at $\sim 1.8 \times 10^{16}$ and final IF is at $\sim 2.7 \times 10^{16}$. The difference between analytical and computational may be due to the  approximate recombination rates used in the estimates, which also require an assumed constant temperature  while calculating the stromgren radius.

\subsection{Warm Absorber Model: Level Populations}
We now explore the time dependent photoionization of warm absorber clouds.   The important free parameters describing warm absorbers are the flux incident on the cloud, the gas density, and the cloud column density.  Table \ref{tab1} shows the parameters spanned by our computations.  As with the H II region, we assume a sudden factor of 3 increases in the illuminating flux at the beginning of the calculation. We illustrate our results using models 73 (baseline model), 74 (high flux high column model), 75 (high density model), and 76 (step down flux model). As shown in Table \ref{tab1}, these span a range in ionization parameter from log($\xi$) = 1.6 to 2.2 and column density from $N_H = $ 0.9 to $2.0\times 10^{23}$ cm$^{-2}$.

A simple description of the time evolution of a warm absorber cloud is as follows:  When the flux of radiation incident on the face of a warm absorber cloud is increased, the ionization rate of the atoms and ions is increased. The increased ionization causes an increase in the photoionization heating of the gas via fast photoelectrons. If the cooling rate is a monotonically increasing function of temperature, then the temperature will increase in order to balance the increased heating. The ionization fractions of the ions  will continue to increase until they reach a balance with the recombination rate. The temperature will continue to increase  until the total heating and cooling rates balance each other, which is the final equilibrium.

\begin{figure*}[ht]
\centering
\includegraphics[width=1\textwidth]{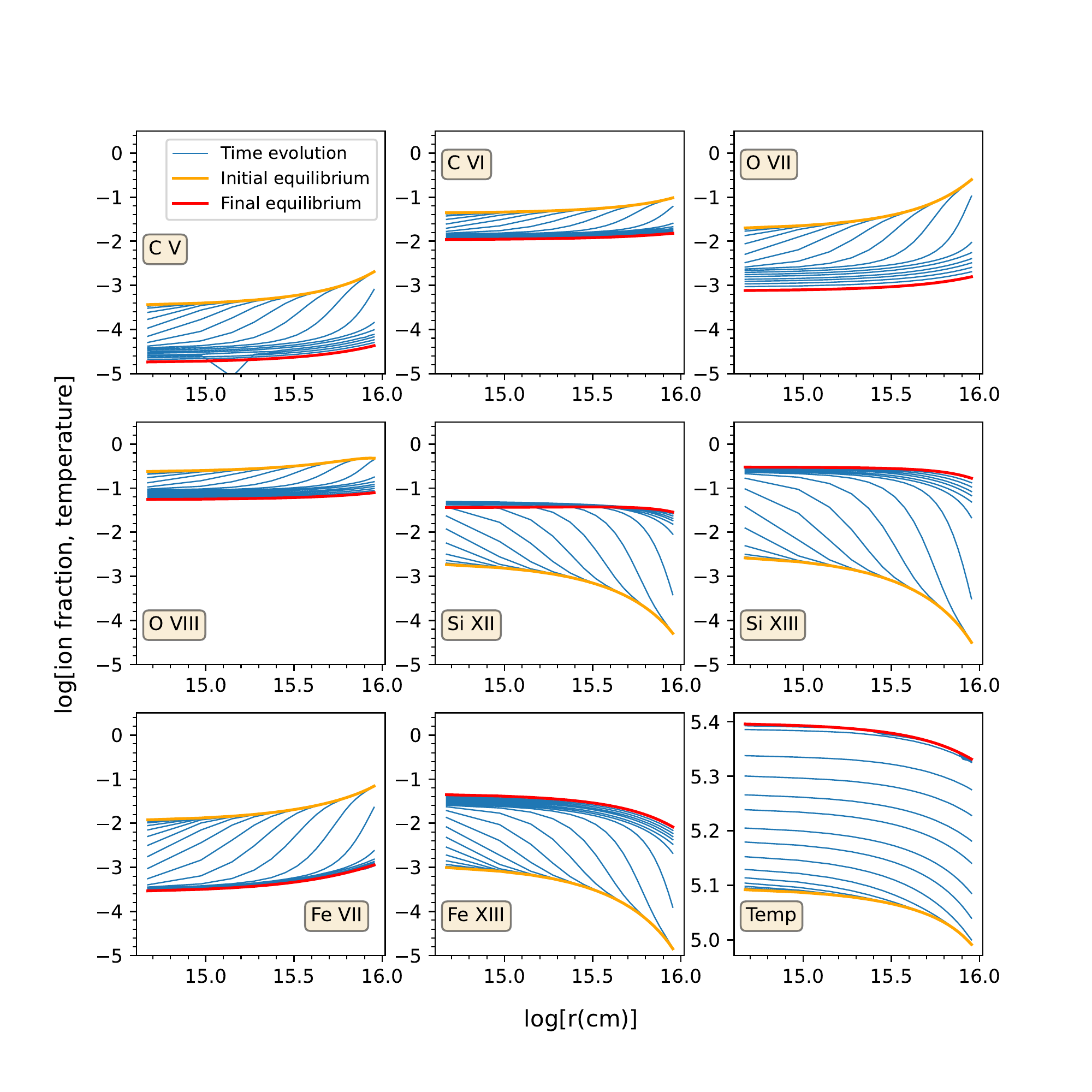}

\caption{Evolution of ion fraction profiles in WA for model 73 (baseline model). The x-axes represent the depths r in the gas from the illuminated face, and the y-axes represent the ion fractions of different ions. The orange curve represents the lower flux equilibrium, while the red curve represents the high flux equilibrium. Each blue curve corresponds to the quantity at an intermediate time during evolution.}

\label{fig4}
\end{figure*}

\begin{figure*}
    \centering
    \includegraphics[width=1\textwidth]{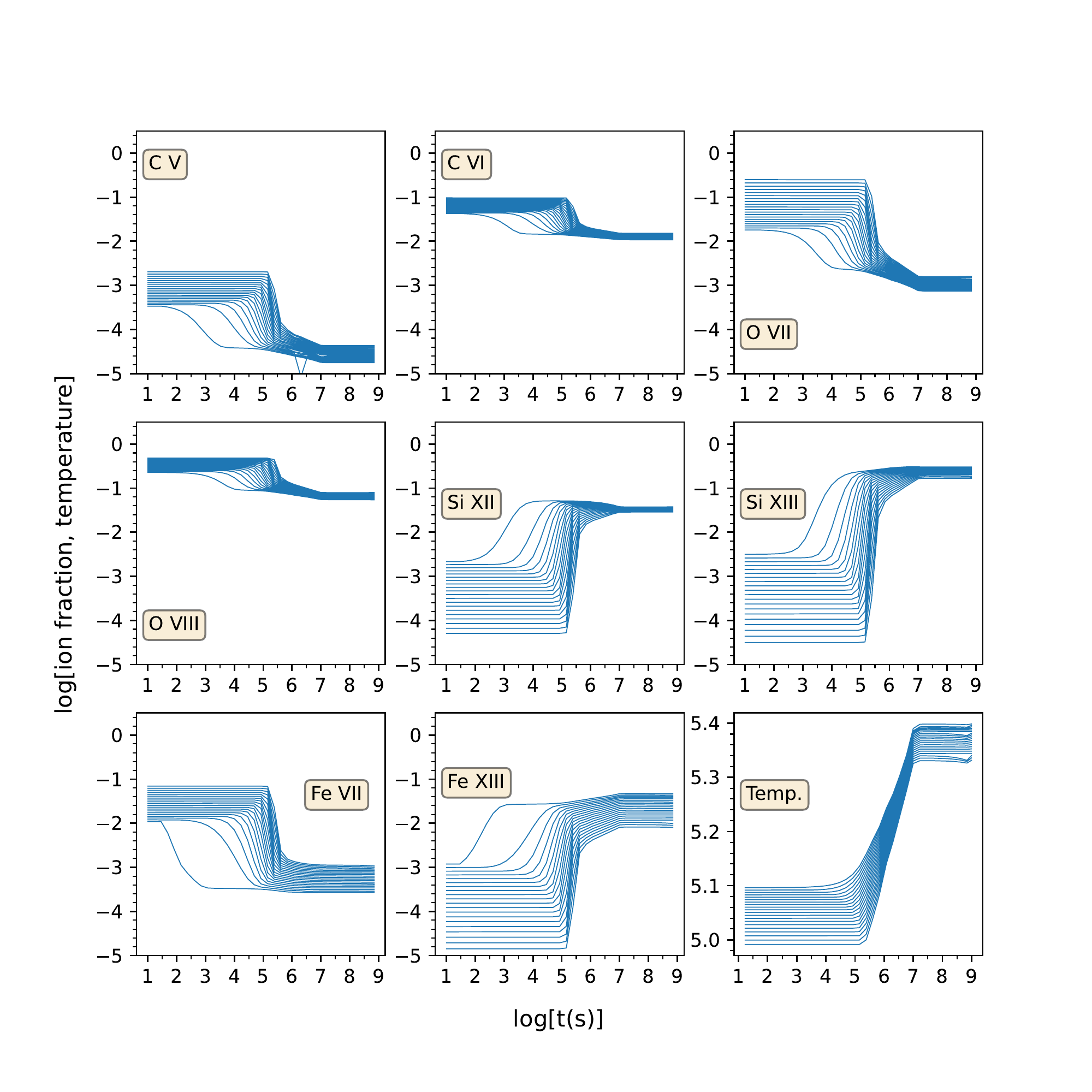}
    \caption{Time evolution of ion fraction for different ions for model 73 (baseline model). Each curve in all panel corresponds to a particular spatial point in the cloud.}
    \label{ionfr_vs_time}
\end{figure*}

Figure \ref{fig4} shows the time evolution of ion fractions and temperature for warm absorber  model 73 (baseline model). All the plots start from the equilibrium values, which are drawn in orange. When we increase the flux, the ionization structure of the cloud starts to change. The changes do not happen instantaneously because of the combined effects of photoionization, propagation, and light travel time. The blue curves correspond to the models at times after the flux increase has occurred; they show the time evolution. The few initial times are 10 s, 17 s, 29 s, 49 s, 83 s, and some final times are $2.4\times 10^8$ s, $4.0\times 10^8$ s, $7.0\times 10^8$ s, $1.0 \times 10^9$ s. These lines are separated by the time $ \Delta {\rm log}(t) \simeq 0.23$. The timescales for this model are approximately as follows:  $t_{pi}\simeq 3.0\times10^3$ s at the face of the cloud, $t_{light}\simeq~3.0\times~10^5$~s, $t_{prop}\simeq 8.2\times 10^5$~s. This corresponds to the yellow shaded region in figure~\ref{fig1}.   

The first panel of figure \ref{fig4} shows the time evolution of the ion fraction of C V. The ion fraction did not change for the first $\sim 3000$ s. This is due to the finite photoionization timescale. It changes first at the face of the cloud. At later times, a wave of increased ionization moves into the cloud. The wave-like lines appear until 3.0 $\times$ 10$^5$ s for model 73 (baseline model). Then the lines start stacking on top of each other. The timescale to reach this point is the propagation timescale.  Subsequently, the shapes of the  lines are similar, although their amplitude continues to change.

This process continues until the gas comes to the final thermal equilibrium given by the final flux. The red line shows this state in the plot. For this model, it takes $\sim 10^7$ s to come to the final thermal equilibrium. This corresponds to the thermal timescale given by equation \ref{thermtime} in section 2.3. The processes responsible for this timescale include electron-ion collision and recombination.  The longer thermal timescale associated with these processes, compared with photoionization and propagation,  is responsible for the late time evolution of the temperature and ion fractions in this model.

In figure \ref{fig4}, the initial equilibrium ion fraction is higher than in the final equilibrium, as shown by the red curve for C V, C VI, O VII, O VIII, and Fe VII. This means that already with the initial flux, the parent atom of these ions was ionized to this state. However, this is not the case for Si XII, Si XIII, and Fe XII. These ions are formed when the flux is increased. The low ionization ions are then converted to these ions, increasing the number of these ions. The bottom right panel shows the temperature evolution in the gas. The orange color represents the initial equilibrium temperature in the gas, which starts to increase after the flux is increased.
Figure \ref{fig4} clearly shows that after the IF has propagated through the cloud, the ion fractions and temperature continue to evolve.  However, this happens on a much longer timescale, the  thermal timescale. This leads to the darker blue band close to the final equilibrium.  In this band, the ion fractions no longer show the wave appearance, and the shape of the ion fractions vs. position is nearly independent of time. 

The time evolution of the ion fractions and temperature can also be visualized by plotting against time. Fig \ref{ionfr_vs_time} illustrates this for model 73 (baseline model). All the panels except the bottom right refer to an ion fraction. The bottom right panel shows the temperature. Individual lines within each panel correspond to the time evolution at a particular spatial point in the cloud. 
To visualize how these quantities change over time, we have taken different spatial points. The point next to the face of the cloud is 4.7$\times 10^{14}$ cm and the rest of the points are separated by $\Delta {\rm log}(R) \simeq 0.3$. 
Ions C V, C VI, O VII, O VIII, and Fe VII initially have higher fractions and decrease at later times. The fractions increase with time for Si XII, Si XIII, and Fe XIII. This model has ionization parameter $\xi \sim$ 100, density of 10$^7$ cm$^{-3}$ and column density of 1.4 $\times$10 $^{23}$ cm$^{-2}$. Thus the size (thickness) of the cloud is $\sim$ 10$^{16}$ cm. The cloud is at 1.8$\times$10$^{17}$ cm from the ionizing source of the initial luminosity of 10$^{44}$ erg s$^{-1}$.

Fig \ref{ionfr_vs_time} shows that at the face of the cloud, the ion fractions start evolving earlier at the photoionization timescale, which is the inverse of the photoionization rate and different for each individual ion after changing flux. Whereas they start to evolve at later times deeper in the cloud at a time which is the  sum of two timescales 1) light travel time (propagation time) and 2) photoionization timescale. The shape of each curve is mainly due to the response of the gas to the radiation field and is characterized by propagation time and photoionization timescale.
 
At late times, photoionization and recombination rates balance, and the ion fraction remains approximately constant. Similar behavior occurs for Si XII, Si XIII, and Fe XIII ions but in the reverse direction. The bottom right panel shows the temperature variation over time at different locations in the cloud. The gas starts heating when it receives the flux to approach another equilibrium temperature. Comparing the ion fraction and temperature plots, the temperature evolves over a longer  timescale ($\sim 10^7$ s) than the photoionization timescale ($\sim 10^4$ s). This behavior is the same as we have seen in figure \ref{fig4} and described in the fourth paragraph of section 4.3. The photoionization time, however, is different for different ions for the flux chosen. All the ions show slight evolution even after photoionization equilibrium because of the change in the temperature. Once the gas attains thermal equilibrium, the ion fraction remains constant.

\subsection{Warm Absorber Spectra}\label{waspec}
\begin{figure*}[ht]
\centering

\includegraphics[width=1.\textwidth]{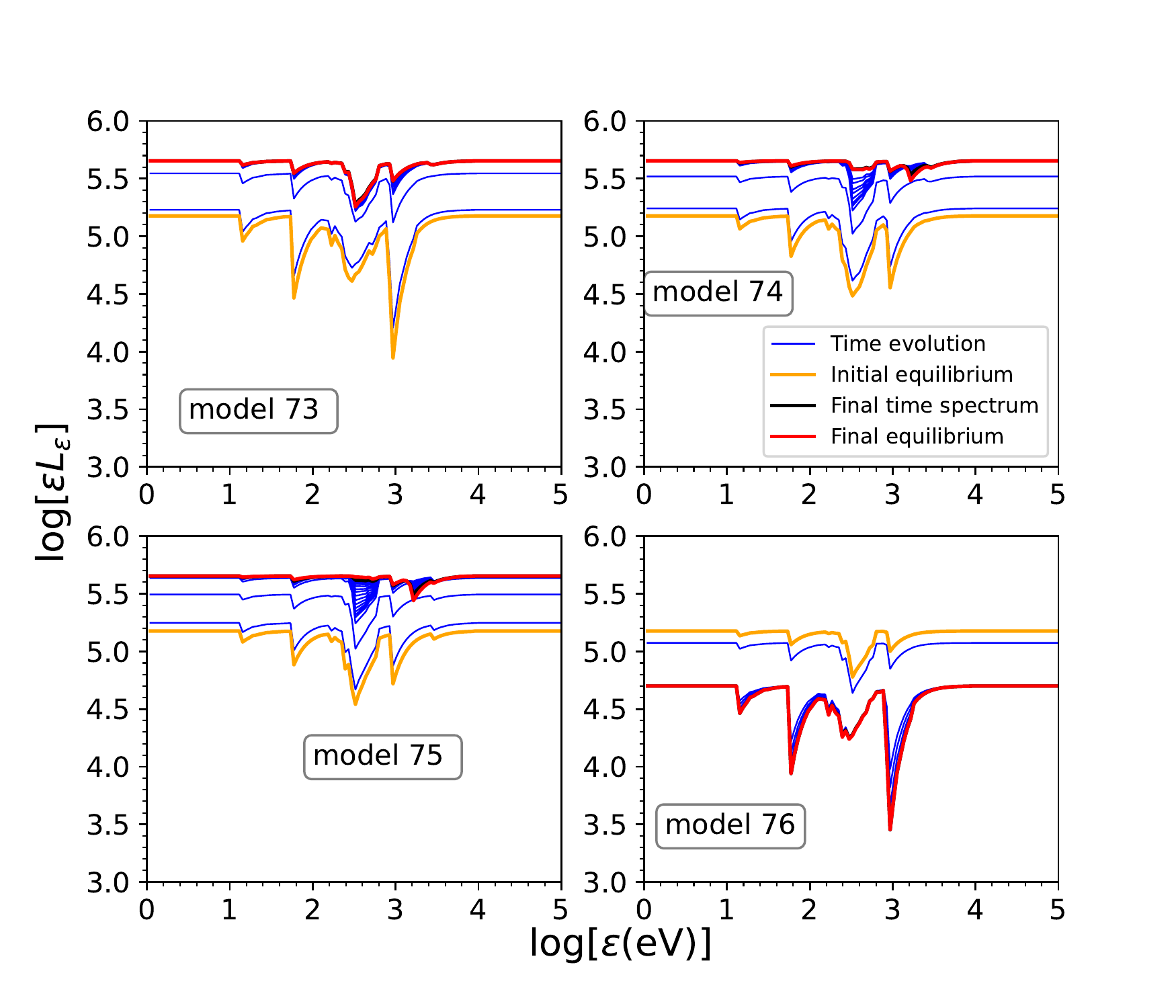}
\caption{Time-dependent WA spectra for four models: 73 (baseline model), 74 (high flux high column model), 75 (high density model), and 76 (step down flux model) for a step incident flux. The orange curves represent the low flux equilibrium spectra, the red curves represent the high flux equilibrium spectra, the blue curves represent the time evolution spectra, and the black color spectrum corresponds to the final time of the simulation. The y-axes represent the transmitted luminosities ($\varepsilon L_\varepsilon$), and the x-axes represent the photon energies ($\varepsilon$). The units of $\varepsilon L_\varepsilon$ are $10^{38}$ erg s$^{-1}$. Model 73 (baseline model), 74 (high flux high column model) and model 76 (step down flux model) were run for $10^9$ s, and model 75 (high density model) for $10^5$ s.}
\label{fig6}
\end{figure*}

Figure \ref{fig6} shows the time evolution of warm absorber spectra for representative models.  The free parameters for those models are listed in table \ref{tab1}.  Model 73 (baseline model), 74 (high flux high column model), and 76 (step down flux model)have the same density of 10$^7$ cm$^{-3}$ but different ionization parameters and column density, while model 75 is a high density model with a density of 10$^{11}$ cm$^{-3}$. These models display four primary absorption features: the lowest is the absorption edge at $\sim 13.6 $ eV, corresponding to hydrogen ionization.  Next is the edge from He II $\sim 54.6$ eV.  The third edge feature, near $\sim$ 328 eV, comes from a combination of absorption from several ions of carbon, silicon, and iron mainly from Fe XII – Fe XV; the fourth edge comes from O VII at $\sim 740$ eV and O VIII at $\sim 870$ eV. In some spectra, the Si K edge is apparent near 2 keV. Models 74 and 75 (high density model) have the same ionization parameters, which are greater than models 73 (baseline model) and 76 (step down flux model), thus giving shallower absorption edges. 
We  have compared equilibrium spectra calculated using {\sc xstar} with equilibrium spectra  calculated using the time dependent code shown in figure 8. These were in good agreement.

As in the previous figures, the orange and red lines correspond to the lower and higher flux equilibrium spectrum, and the black curve corresponds to the last time point.  The blue lines represent the spectrum at intermediate times during the evolution. Since we have logarithmic time grids, we can see only two distinct intermediate spectra. All others overlap at either earlier or later times at all energies except at the third edge. The first intermediate blue spectrum close to the orange spectrum is at time $\sim 2.4 \times 10^5$ s, 
and another one near equilibrium (red) spectrum at $\sim 4.1 \times 10^5$ s for models 73 (baseline model) and 74 (high flux high column model). The spectra before this time overlap the initial equilibrium curve because of the light travel time plus characteristic photoionization time. The evolution timescale, however, is shorter for the high density ($10^{11}$ cm$^{-3}$) model 75 (high density model). The two distinct intermediate blue spectra are at 52 s and 66 s. One of the two distinct intermediate spectra close to the orange color spectrum in model 76 (step down flux model) is at time $\sim 2.4 \times 10^5$ s, and one close to the red spectrum is at$\sim 4.1 \times 10^5$ s. The prominent blue band in the $\sim$328 eV edge in models 74 (high flux high column model) and 75 (high density model) is due to the slower evolution of the iron ions in Fe XII --  Fe XV. The reason for the slower evolution of the edges in this energy band is because of the several closely spaced ionization potentials of the ions of the iron.  Each of these has its own photoionization timescale, which imposes an added delay in the response of the gas. The stacking of blue lines at $\simeq$ 328 eV edge comes at time  $\sim 4.1 \times 10^5$ s and later in model 74 (high flux high column model)and 66 s or later in model 75 (high density model).

The spectrum consists of distinct spectra. There are two different effects causing the spectrum to change in figure 7. One is in the continuum, which is not affected by absorption. Ideally, the sudden change in the illumination should correspond to a sudden change in the transmitted spectrum in the parts of the spectrum which are not affected by absorption. In our models, the transition has been smeared due to finite time resolution and numerics. This effect is minor for a smaller time and more significant for a larger time, as seen in the figure, and can be minimized by taking more time grid points.

The second important effect that changes the spectrum is due to the time dependence of the optical thickness. When the gas at different depths experiences a sudden rise in flux, it starts to photoionize the gas and becomes more transparent afterward. More light can be transmitted through the gas, causing variability in the absorption edges.


The convergence of the spectrum is mainly guided by the photoionization time. Once the gas reaches the photoionization equilibrium, the spectrum comes close to the final equilibrium spectrum. The small change comes from the evolution of ion fractions due to temperature evolution. The reason for the slower temperature evolution is described  in the previous subsection. All the models converge with the final flux equilibrium spectrum.

\begin{figure*}[ht]
\centering
\includegraphics[width=0.8\textwidth]{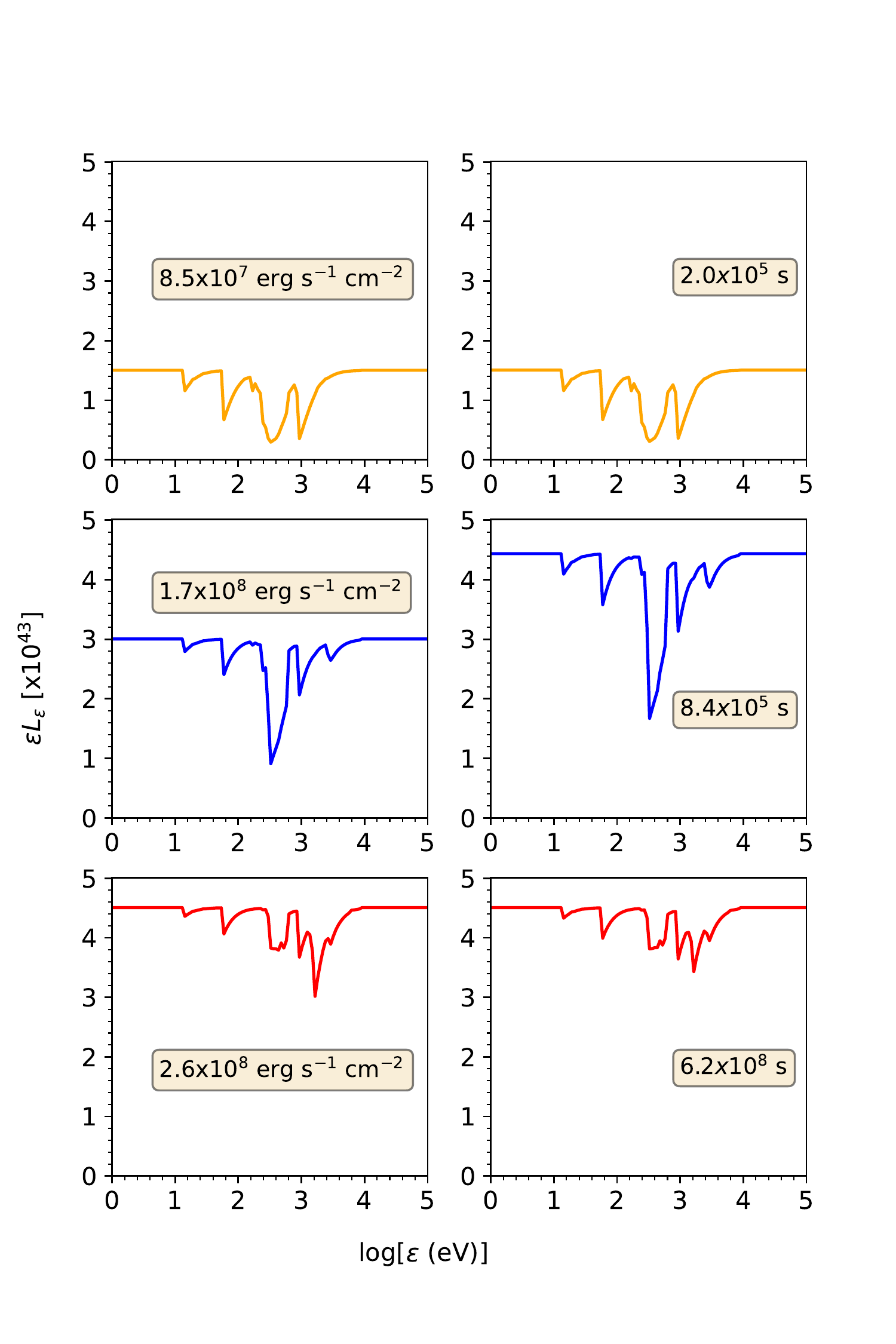}
\caption{Comparison of equilibrium spectra with time dependent spectra for model 74 (high flux high column model). The left column shows the spectra with absorption edges for different incident ionizing fluxes considering the equilibrium state case. The right panel is an instantaneous spectrum obtained at three different times during its evolution. The units of $\varepsilon L_\varepsilon$ are erg~s$^{-1}$ in y-axes.}
\label{fig5}
\end{figure*}
 
Figure \ref{fig5} illustrates how the equilibrium spectra differ from the spectra considering time dependent calculation for model 74 (high flux high column model). The left column shows spectra from  equilibrium calculations for various fluxes:  the initial ionizing flux ($F_{ion,1}$) $8.5 \times 10^7$ erg s$^{-1}$ cm $^{-2}$, the final ionizing flux ($F_{ion,2}$) ($2.6~\times~10^8$~erg~s$^{-1}$~cm~$^{-2}$) and the average of these two ($1.7~\times 10^8$~ erg~s$^{-1}$~cm~$^{-2}$). The column on the right comes from the time dependent calculation model 74 (high flux high column model) and shows representative spectra at three different times; $5.5 \times 10^5$~s, $8.9 \times 10^6$~s, and $6.2 \times 10^8$~s, respectively, from the top. Looking at these equilibria and time dependent spectra, it is seen that they are different from one another before they converge to the final equilibrium spectrum.
Even though the shape looks somewhat similar between the second panel of both columns, the depth of the absorption edges and continuum are significantly different. This is a clear indication of a need for a time dependent calculation for these conditions. It shows that fitting a family of equilibrium model spectra to observations of a naturally time-varying warm absorber will not lead to fitting parameters which are accurate.  Fitting parameters such as ionization parameters and element abundances may be significantly in error.

Figure \ref{fig5} is  designed to show that time dependent spectra differ from equilibrium. Since we are working with an incident light curve which undergoes a sudden step up or step down, there is no intermediate incident flux value. For comparison, the equilibrium model shown on the left is run for an average of the initial and final incident flux values. This kind of comparison is helpful to see the difference when we try to approximate the true time dependent spectra with time-average spectra. Since the second panels of both columns have different continuum levels, the optical depths at the energy around 328 eV are calculated. The optical depths for the first and last panels are the same and have values of 2.01 and 0.16, respectively. While for middle panels, $\tau \approx 1.32$ for equilibrium spectra and $\tau \approx 1.03$ for time dependent spectra. This demonstrates that the time dependent spectra are changing even after the ionizing flux becomes constant. This is clearly different from what the equilibrium approximation predicts. 

\begin{table*}[]
\centering
\begin{tabular}{|c|llllllllllllllll|}
\hline
Energy (eV) & \multicolumn{16}{c|}{Model} \\ \hline
 & \multicolumn{1}{l|}{\textbf{13}} & \multicolumn{1}{l|}{\textbf{33}} & \multicolumn{1}{l|}{\textbf{73}} & \multicolumn{1}{l|}{\textbf{93}} & \multicolumn{1}{l|}{\textbf{14}} & \multicolumn{1}{l|}{\textbf{34}} & \multicolumn{1}{l|}{\textbf{74}} & \multicolumn{1}{l|}{\textbf{94}} & \multicolumn{1}{l|}{\textbf{15}} & \multicolumn{1}{l|}{\textbf{35}} & \multicolumn{1}{l|}{\textbf{75}} & \multicolumn{1}{l|}{\textbf{95}} & \multicolumn{1}{l|}{\textbf{16}} & \multicolumn{1}{l|}{\textbf{36}} & \multicolumn{1}{l|}{\textbf{76}} & \textbf{96} \\ \hline
 & \multicolumn{16}{c|}{log[response time($t_{1/2}$]) in s} \\ \hline
14.3 & \multicolumn{1}{l|}{5.5} & \multicolumn{1}{l|}{5.7} & \multicolumn{1}{l|}{5.4} & \multicolumn{1}{l|}{5.6} & \multicolumn{1}{l|}{5.7} & \multicolumn{1}{l|}{5.9} & \multicolumn{1}{l|}{5.6} & \multicolumn{1}{l|}{5.9} & \multicolumn{1}{l|}{1.7} & \multicolumn{1}{l|}{1.9} & \multicolumn{1}{l|}{1.7} & \multicolumn{1}{l|}{1.9} & \multicolumn{1}{l|}{5.6} & \multicolumn{1}{l|}{5.7} & \multicolumn{1}{l|}{5.7} & 5.6 \\ \hline
59.3 & \multicolumn{1}{l|}{5.5} & \multicolumn{1}{l|}{5.8} & \multicolumn{1}{l|}{5.5} & \multicolumn{1}{l|}{5.7} & \multicolumn{1}{l|}{5.7} & \multicolumn{1}{l|}{6.0} & \multicolumn{1}{l|}{5.7} & \multicolumn{1}{l|}{5.9} & \multicolumn{1}{l|}{1.8} & \multicolumn{1}{l|}{2.0} & \multicolumn{1}{l|}{1.8} & \multicolumn{1}{l|}{1.9} & \multicolumn{1}{l|}{5.5} & \multicolumn{1}{l|}{5.6} & \multicolumn{1}{l|}{5.6} & 5.5 \\ \hline
271 & \multicolumn{1}{l|}{5.5} & \multicolumn{1}{l|}{5.8} & \multicolumn{1}{l|}{5.5} & \multicolumn{1}{l|}{5.7} & \multicolumn{1}{l|}{5.7} & \multicolumn{1}{l|}{6.0} & \multicolumn{1}{l|}{5.7} & \multicolumn{1}{l|}{6.0} & \multicolumn{1}{l|}{1.8} & \multicolumn{1}{l|}{2.0} & \multicolumn{1}{l|}{1.8} & \multicolumn{1}{l|}{1.9} & \multicolumn{1}{l|}{5.5} & \multicolumn{1}{l|}{5.6} & \multicolumn{1}{l|}{5.6} & 5.5 \\ \hline
328 & \multicolumn{1}{l|}{5.4} & \multicolumn{1}{l|}{5.8} & \multicolumn{1}{l|}{5.5} & \multicolumn{1}{l|}{5.7} & \multicolumn{1}{l|}{7.2} & \multicolumn{1}{l|}{7.9} & \multicolumn{1}{l|}{6.3} & \multicolumn{1}{l|}{6.8} & \multicolumn{1}{l|}{3.0} & \multicolumn{1}{l|}{3.7} & \multicolumn{1}{l|}{2.2} & \multicolumn{1}{l|}{2.7} & \multicolumn{1}{l|}{5.6} & \multicolumn{1}{l|}{5.7} & \multicolumn{1}{l|}{5.7} & 5.6 \\ \hline
436 & \multicolumn{1}{l|}{5.4} & \multicolumn{1}{l|}{5.7} & \multicolumn{1}{l|}{5.5} & \multicolumn{1}{l|}{5.6} & \multicolumn{1}{l|}{6.4} & \multicolumn{1}{l|}{7.5} & \multicolumn{1}{l|}{5.8} & \multicolumn{1}{l|}{6.6} & \multicolumn{1}{l|}{2.0} & \multicolumn{1}{l|}{3.3} & \multicolumn{1}{l|}{1.9} & \multicolumn{1}{l|}{2.5} & \multicolumn{1}{l|}{5.6} & \multicolumn{1}{l|}{5.7} & \multicolumn{1}{l|}{5.7} & 5.6 \\ \hline
579 & \multicolumn{1}{l|}{5.4} & \multicolumn{1}{l|}{5.7} & \multicolumn{1}{l|}{5.4} & \multicolumn{1}{l|}{5.6} & \multicolumn{1}{l|}{5.7} & \multicolumn{1}{l|}{6.0} & \multicolumn{1}{l|}{5.7} & \multicolumn{1}{l|}{6.0} & \multicolumn{1}{l|}{1.8} & \multicolumn{1}{l|}{2.0} & \multicolumn{1}{l|}{1.8} & \multicolumn{1}{l|}{2.0} & \multicolumn{1}{l|}{5.6} & \multicolumn{1}{l|}{5.7} & \multicolumn{1}{l|}{5.7} & 5.6 \\ \hline
770 & \multicolumn{1}{l|}{5.4} & \multicolumn{1}{l|}{5.7} & \multicolumn{1}{l|}{5.4} & \multicolumn{1}{l|}{5.6} & \multicolumn{1}{l|}{5.6} & \multicolumn{1}{l|}{5.9} & \multicolumn{1}{l|}{5.6} & \multicolumn{1}{l|}{5.9} & \multicolumn{1}{l|}{1.7} & \multicolumn{1}{l|}{1.8} & \multicolumn{1}{l|}{1.7} & \multicolumn{1}{l|}{1.8} & \multicolumn{1}{l|}{5.6} & \multicolumn{1}{l|}{5.7} & \multicolumn{1}{l|}{5.7} & 5.6 \\ \hline
847 & \multicolumn{1}{l|}{5.6} & \multicolumn{1}{l|}{5.8} & \multicolumn{1}{l|}{5.5} & \multicolumn{1}{l|}{5.7} & \multicolumn{1}{l|}{5.7} & \multicolumn{1}{l|}{5.9} & \multicolumn{1}{l|}{5.6} & \multicolumn{1}{l|}{5.9} & \multicolumn{1}{l|}{1.7} & \multicolumn{1}{l|}{1.8} & \multicolumn{1}{l|}{1.7} & \multicolumn{1}{l|}{1.8} & \multicolumn{1}{l|}{5.6} & \multicolumn{1}{l|}{5.7} & \multicolumn{1}{l|}{5.7} & 5.6 \\ \hline
931 & \multicolumn{1}{l|}{5.6} & \multicolumn{1}{l|}{5.9} & \multicolumn{1}{l|}{5.6} & \multicolumn{1}{l|}{5.8} & \multicolumn{1}{l|}{5.8} & \multicolumn{1}{l|}{6.0} & \multicolumn{1}{l|}{5.7} & \multicolumn{1}{l|}{6.0} & \multicolumn{1}{l|}{1.8} & \multicolumn{1}{l|}{2.0} & \multicolumn{1}{l|}{1.8} & \multicolumn{1}{l|}{2.0} & \multicolumn{1}{l|}{5.5} & \multicolumn{1}{l|}{5.6} & \multicolumn{1}{l|}{5.6} & 5.5 \\ \hline
1020 & \multicolumn{1}{l|}{5.6} & \multicolumn{1}{l|}{5.9} & \multicolumn{1}{l|}{5.6} & \multicolumn{1}{l|}{5.7} & \multicolumn{1}{l|}{5.7} & \multicolumn{1}{l|}{6.0} & \multicolumn{1}{l|}{5.7} & \multicolumn{1}{l|}{6.0} & \multicolumn{1}{l|}{1.8} & \multicolumn{1}{l|}{2.0} & \multicolumn{1}{l|}{1.8} & \multicolumn{1}{l|}{1.9} & \multicolumn{1}{l|}{5.5} & \multicolumn{1}{l|}{5.6} & \multicolumn{1}{l|}{5.6} & 5.5 \\ \hline
\end{tabular}
\caption{Response time ($t_{1/2}$) in s for different energy values for different models. The bold face numbers are the model name.}
\label{tab2}
\end{table*}

\begin{figure*}[ht]
\centering
\includegraphics[width=0.9\textwidth]{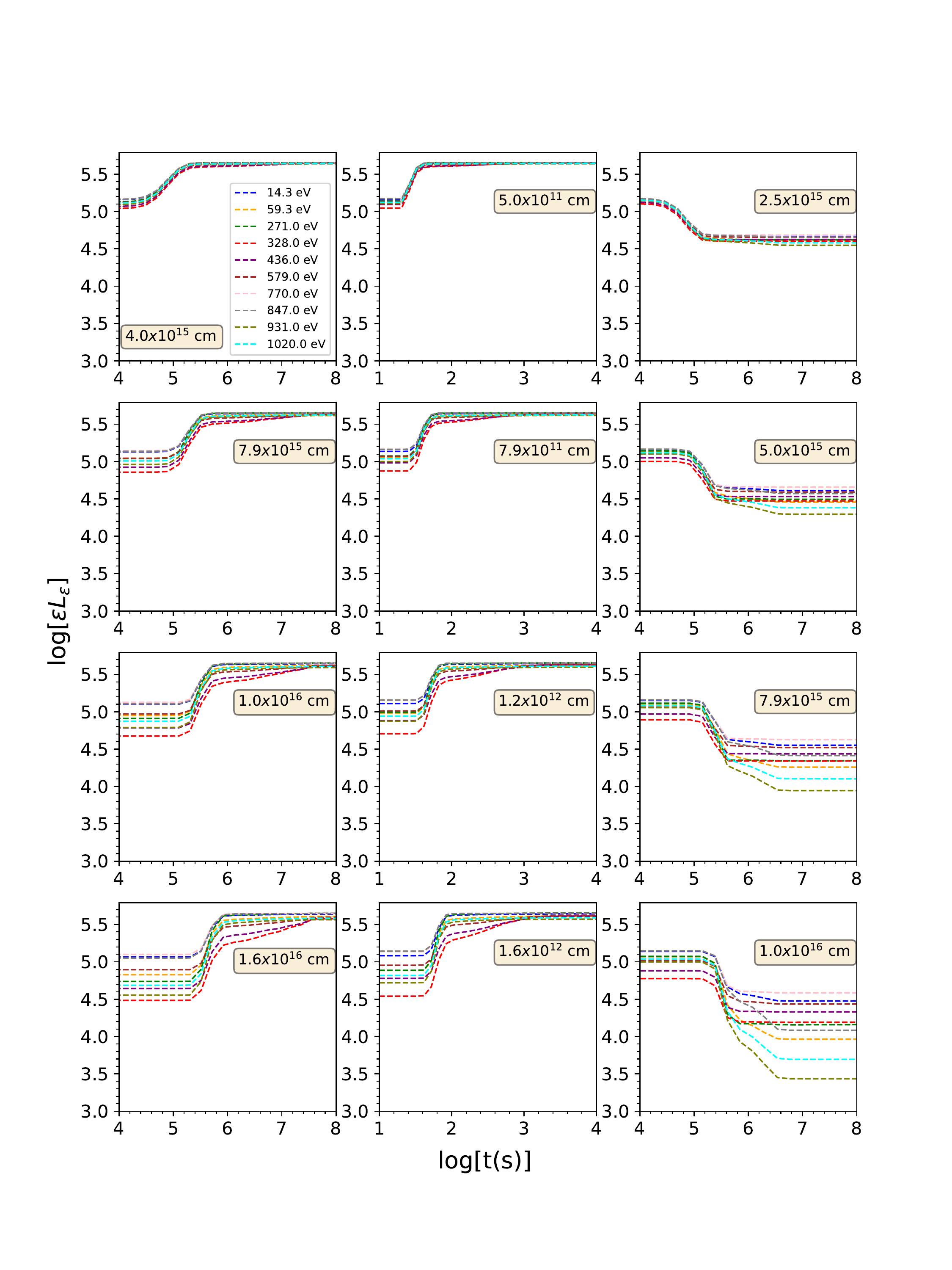}
\caption{Time variation of specific luminosity for ten different energies at different depths as transmitted within the cloud noted by value in cm appearing in the box. The left panel shows the time evolution for model 74 (high flux high column model), the middle for model 75 (high density model), and the right for model 76 (step down flux model). The units of $\varepsilon L_\varepsilon$ are $10^{38}$ erg s$^{-1}$ in y-axes. X-axes represent the t in s. The time range shown here is part of the total time of the simulation. The simulation starts at 10 s, and the initial flux change occurs at 20 s.}
\label{fig7}
\end{figure*}
\subsection{Warm Absorber Light Curves}
More detail about the behavior of the transmitted luminosity is shown in figure \ref{fig7}. This shows the time variation of the specific luminosity at particular photon energies at different positions in the cloud for models 74 (high flux high column model), 75 (high density model), and 76 (step down flux model) for the case where the flux in changed suddenly at the face of the cloud.  The times in this figure are relative to the time when the jump occurs at the face of the cloud (not at the source). This figure includes three columns for three models, and each column contains four panels. The y-axis is the log of the product of specific luminosity and energy, $\varepsilon L_{\varepsilon}$. The x-axis is a log of time, and the range is different for model 75 from the other two. The details of each panel are described below. 

The face of the cloud is at $\sim$ 3.0 $\times 10^{17}$ cm from the central ionizing source of active galactic nuclei for model 74 (high flux high column model). The size of the cloud for this model is $1.6 \times 10^{16}$ cm. When we move to $4.0 \times 10^{15}$ cm from the illuminating face,
we see the variation as shown in panel 1. Different lines correspond to different energy values. For example, the blue dash curve represents 14.2 eV, which is close to the ionization threshold of the hydrogen atom. The plot shows that it starts to evolve only at $\sim 10^5$ s due to the light travel time from the cloud face. The absorption differs for different energy values due to the fact that the absorption cross-section is  energy-dependent. The photons at energy near $\sim 328$ eV are absorbed more efficiently by the gas compared to the H Lyman continuum photons. The flat part of the curve represents the low flux equilibrium state of the cloud.  The slope represents how the optical depth is changing in the gas between this point and the face of the cloud. The ionization stage of the gas starts to change and become transparent for some energies, while it may be opaque for other energies. The rate of change of optical depth, however, is different for different photon energy. If we let the gas evolve for a long time, the specific luminosity value converges to the higher flux equilibrium.

The time evolution of the specific luminosity at positions $\sim 7.9 \times 10^{15}$ cm, $\sim 1.0 \times 10^{16}$ cm, $\sim 1.6 \times 10^{16}$ cm are shown in the second, third, and fourth panels respectively. Even though the shapes of the curves look similar, they are different in their  transition from the initial to the final equilibrium value, amount of absorption, and response time. If we define the response time as the time it takes to reach  approximately half of the equilibrium value, then the second panel shows it takes $\sim2.3 \times 10^5$ s to reach the depth of $\sim 7.2 \times 10^{15}$ cm, and hence the luminosity starts to increase at later times than in the top panel 1. In the same way, it takes $\sim 3.3 \times 10^5$ s in the third panel and $\sim 5.3 \times 10^5$ s in the fourth panel. These differences are due to light travel time. The amount of absorption also increases in the subsequent panel as column density increases if we go deeper into the cloud. Each plot at a different depth is the result of the time evolution effects of the gas in front of it. At early times, the gas is not fully ionized, so some energies are more absorbed than others. At the final times, the gas is nearly fully ionized and transparent for depths less than the full cloud thickness, so fluxes at all energies converge to nearly the same value.

The second column of fig \ref{fig7} shows similar behavior as in model 74 (high flux high column model), but with a greatly differing timescale. Model 75 (high density model) has gas density of $10^{11}$ cm $^{-3}$, ionization parameter $\xi \sim 100$, and source luminosity of $10^{44}$ erg s$^{-1}$. These parameter sets the gas dimension to be $\sim 1.6 \times 10^{12}$ cm, which is much thinner than the other models. The cloud, in this case, is at  $\sim 10^{15}$ cm from the central ionizing source. The evolution pattern of light curves at all depths is similar to model 74. The only difference is the light travel time or the propagation time, which is $\sim$ 17 s for the first panel and $\sim$ 53 s for the last panel. It takes to go from a lower flux equilibrium value to a higher one is about 1000 s. If we compare this timescale with model 74, we see that Model 75 has a higher gas density by an order of 10$^4$ and hence the response time is smaller by a comparable factor.

The third column of figure \ref{fig7} is for model 76, which corresponds to the step down model. Free parameters are given in table \ref{tab1}. This sets the cloud size to be $\sim 10^{16}$ cm. The gas is initially exposed to a high flux. When we lower the flux by a faction of 3 suddenly, the gas starts to recombine, and gas becomes optically thicker over time. Hence the transmitted luminosity decreases as time passes. For a step down model, the  time required to reach equilibrium is fixed by the recombination timescale and cooling timescale, which is longer than the photoionization  timescale and heating timescale. Hence this model is still evolving at some energies even after $10^8$ s. 

A convenient way to describe the response of a model cloud to a sudden change in the flux is to  define a quantity we call response time. This is defined as the time it takes for a quantity to  reach  a value halfway between the initial and final equilibrium values in the light curve. Here, we focus on the response time for the specific luminosity. This provides a convenient way of describing how the light curve at a given energy and location in the cloud evolves from lower equilibrium to higher equilibrium. Table \ref{tab2} gives these timescales for the 16 warm absorber models defined in table \ref{tab1}. We find the response time at all energy values for models 13, 33, 73 (step down flux model), and 93 to be approximately the same. However, the response time at energy $\sim 328$ eV and $\sim 436$ eV for models 14, 34, 74 (high flux high column model), and 94 are significantly longer than the values at other energies, as discussed in section 4.3.  
The response time ($t_{1/2}$) for model 74 is $\sim 2.0 \times 10^{6}$ s for $\sim$ 328 eV and $\sim 6.3 \times 10^{5}$ s for $\sim$ 436 eV in comparison with $\sim 4.4 \times 10^{5}$ s for $\sim$ 14 eV and $\sim 5.5 \times 10^{5}$ s for $\sim$~931~eV. 

The light curve shapes depend on photon energy.  The curve at energies such as 14.3 eV and 59.3 eV evolve quickly and hence have higher slope  $\sim 3 \times 10^{37}$ erg s$^{-2}$. In comparison, the light curve at energies 328 eV and 436 eV, the slopes are much smaller, $\sim 9 \times 10^{35}$ erg s$^{-2}$. The response time  ($t_{1/2}$) in table \ref{tab2} provides another measure of the slope;  a longer response time  generally corresponds to a smaller slope. These timescales are set mainly by the photoionization and recombination time of the absorbing species. The slower evolution at energies 328~eV and 436 eV are described in the previous section \ref{waspec}. 

We have performed simulations of how the gas responds to a sudden change in the ionizing flux. The results of this work are applicable to any change in the flux, which is much faster than the response time of the gas. If the timescale of change in the flux is much slower than the response time of the gas, then the gas will remain in the equilibrium state appropriate to the flux at any given time. If the variability timescale of the input flux is comparable to the response time of the gas, the effect requires a special treatment which we will explore in future work.

\subsection{Power Spectrum}
A convenient  way of analyzing time dependent phenomena is via the Fourier transform. We have performed the Fourier analysis of specific luminosity as a function of time at different energy values. The equation defining the Fourier transform of a function $f(t)$ is given by
\begin{equation}
    f(\nu)=\frac{1}{2 \pi}\int_{-\infty}^ \infty f(t) e^{-i\omega t} dt
\end{equation}
In our case, $f(t)$ and $f(\nu)$ would be $L_{\varepsilon}(t)$ and $L_{\varepsilon}(\nu)$. Where $L_{\varepsilon}(t)$ is the luminosity at energy value $\varepsilon$ as a function of time at different cloud depths in the emitted spectrum. 
 
\begin{figure}[ht]
\centering
\includegraphics[width=0.45\textwidth]{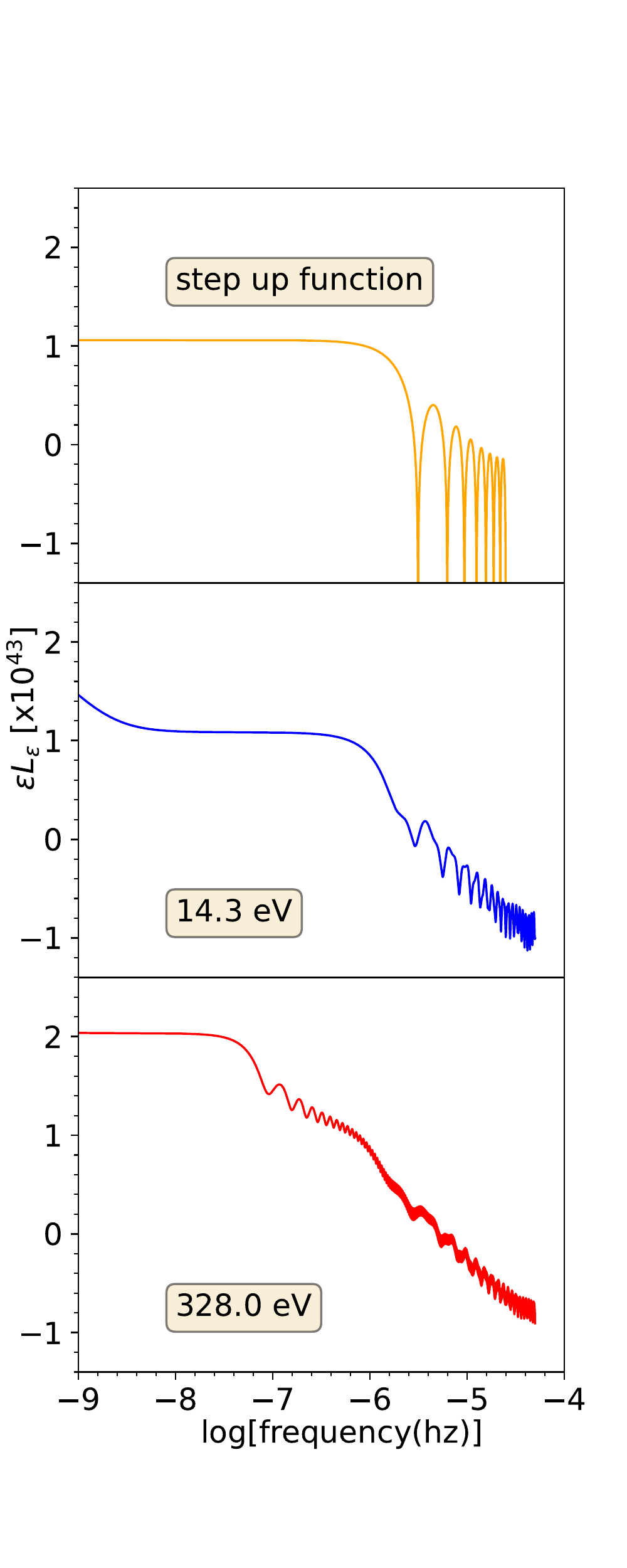}
\caption{Fourier transform of the specific luminosity at different photon energies for model 74 (high flux high column model). The units of $\varepsilon L_\varepsilon$ are erg s$^{-1}$ in y-axes.}
\label{fig9}
\end{figure}

Figure \ref{fig9} shows the Fourier transform of some of the light curves of the last panel of  the first column of figure \ref{fig7}. The  highest frequency  parts of  figure \ref{fig9} are affected by insufficient sampling in time  and show which frequency carries most of the power in the spectrum. For $\sim$ 14 eV, most of the power is in relatively high frequencies meaning the flat region extends to $\sim 10^{-5}$ Hz with amplitude $\sim$1. However, for $\sim$ 327 eV, most of the power is in low frequencies, mostly in $< 10^{-7}$ Hz.

Fig \ref{fig9} shows that time dependent effects introduce a 'knee' in the curve at the characteristic timescales associated with atomic processes in the power spectrum of variability in the warm absorber edge features.

A warm absorber will behave as a filter when applied to the variable signal from the central source.  The central source in most AGN varies on a broad range of timescales with an approximate power law distribution in the power density spectrum \citep{ulr97, utt04}. If the flux from the central source varies slowly, then the warm absorber gas has enough time to respond to the changes. Hence, the time variability of the warm absorber response will be similar to the time variability of the flux from the central source. If the variability timescale of the central source is short in comparison to the response time, the warm absorber will not have enough time to adjust the changes.  The warm absorber variability will then represent a kind of average over the variability of the central source flux.  This response behavior of the warm absorber and its imprint on the time variability of the central source can be detected by comparing observed warm absorber power spectra with synthetic power spectra. Such synthetic spectra can be constructed by constructing the 'transfer function' \citep{edel88}, which is the ratio of the power spectrum of model warm absorbers to the power spectrum of the assumed illuminating radiation light curve. This can then be convolved with real AGN continuum power spectra to get the synthetic warm absorber power spectra.  We plan to carry out such calculations in a subsequent paper.  

\section{Discussion}
In astrophysical plasmas, the timescale for the gas to respond to changing flux can vary over a wide range.  The response timescale primarily depends upon the density of the gas, the radiation field, and the temperature of the gas.  We can have situations ranging from extremely low-density regions like the interstellar medium, having $\sim$few atoms per cubic centimeter,  to highly dense regions like accretion discs in active galactic nuclei,  which may have $\sim 10^{23}$ cm$^{-3}$.  The corresponding response timescales can vary accordingly.   In the case of the warm absorber, timescales for atomic processes like photoionization and recombination and the propagation time can be longer than the typical AGN variability timescales.
Furthermore, even if atomic timescales are short compared with the AGN variability timescale, the propagation time can be longer. This occurs when the supply rate of ionizing photons is relatively small. In this case, we need to take time dependent calculation into consideration.

Ionization and recombination rates for some ions have complicated dependence on the temperature, and the temperature has a non-linear dependence on the radiation field. This has a peculiar impact on the time evolution of gas when the flux is changed. The evolution is not monotonic and is not the same for all photon energies in the spectrum. This led us to define the response time in our calculation which is given in table \ref{tab2}.

Time-resolved spectra can be used to constrain the density of the absorber gas. This procedure has been outlined by \citet{nic99}. For example, for a variable source in which the continuum makes a sudden step up or down, we can observe the spectra for a range of times. Then we can make a light curve in the absorption optical depth as a function of time after the sudden change in the flux. This light curve can then be compared with the results of models at various densities similar to what is shown in table 3.  Models with high densities will have optical depths which can vary rapidly and come to equilibrium quickly; models with low densities will have optical depths which vary more slowly due to the longer timescales for ionization, recombination, and propagation.  Models with some intermediate gas densities will show delayed responses matching the observations. Looking at these responses, we can fit the simulated transmitted spectra to the observation to infer the gas density.

The traditional way of fitting and analyzing spectra, including warm absorber spectra, assumes the emitting or absorbing gas is in equilibrium. Such models have free parameters, including ionization parameter ($\xi$) and column density($N_H$). These parameters are tuned until we get the best fit with the observed spectra. The velocity of outflowing gas can be measured, assuming that spectral lines are broadened by Doppler broadening. This, combined with the virial theorem, can be used to estimate the location of the gas using the relation $R=GM/v^2$. For a  time-varying radiation field, time average flux($F_{ion}$) is calculated, and from there, the luminosity of the AGN ($L_{ion}$) is determined using $L_{ion}=4\pi R^2 F_{ion}$. Finally, one can estimate the density of the gas ($n_H$) using the relation $\xi=L_{ion}/(n_HR^2)$. Once we know the location, density, and speed of the outflow, the outflow rate can be calculated using $\Dot{M}_{wind}=4\pi R^2 v m_H$. In many cases, this leads to a very high mass flow rate,  greater than the accretion rate $\Dot{M}_{accretion}=L_{ion}/(\eta c^2)$, where $\eta$ is the efficiency of converting acreeting materials into electromagnetic energy. This is surprising and can have significant implications for galaxy feedback.

Our calculations show that time dependent spectra are different from corresponding equilibrium spectra qualitatively, as shown in fig \ref{fig6}. The change in the gas is not instantaneous because of the propagation, light travel time,  and photoionization timescales. Some ions and associated edges evolve much more slowly than others due to small photoionization rates, e.g., the iron  edges near $\sim$ 300 - 400 eV. Also,  it is seen that the gas evolves more slowly if the input flux is stepped down rather than if the flux is stepped up. This indicates that a stochastically varying AGN will produce a warm absorber that spends more time looking like an equilibrium model closer to the high flux state rather than the low flux state. In such a situation, the true flux incident on the warm absorber will be less than would be inferred by fitting the sequence of equilibrium models. Using this flux value, we end up with a mass flow rate smaller than the one predicted by equilibrium calculation.

Limitations of our models originate from the fact that  our time dependent computation is time-consuming, so we have simplified the problem by assuming a simple three-level atom or ion as discussed in subsection \ref{subsec3.1}. This reduces the number of equations to be solved significantly. For the same reason, we include  only the elements hydrogen, helium, carbon, oxygen, silicon, and iron and use a relatively small number of spatial grid points, time grid points, and energy grid points while discretizing.  

We model the atom or ion with only three energy levels and take collisional excitation and de-excitation into account for only hydrogen and helium. This causes a limited pathway to process the energy in terms of collisional excitation and decay radiatively, leading collisional cooling to be somewhat underestimated and, therefore, possibly overestimating the temperature of the gas. We also do not predict  absorption lines because of our coarse energy grid. Our energy bins are broader than the width of the line. 

We have not included emissions in our radiative transfer calculation. This approximation will be justified if the covering fraction of the absorber relative to the central source is small; this appears to be justified for most observed warm absorbers. For most observed warm absorber spectra, emission features are weak or absent  \citep{kas01,kas02}.

In this paper, we assume a sudden change in the ionizing flux by a factor of 3. This is an adequate representation of a situation where the ionizing source is changing its luminosity on a very short  timescale.  A flaring incident light curve, in which the flux increases and then decreases back to the initial value, may be closer to reality and give more accurate fits to observations. We will consider this in future work.

\section{Conclusion}
\begin{itemize}

\item We have performed calculations of time dependent photoionization in two different cases corresponding to conditions resembling HII regions  and those resembling AGN warm absorbers. For this, we have solved coupled time dependent differential equations of level population, heating and cooling, and radiative transfer. Our simulations show that the time dependent effects are important and different from the steady-state solution in terms of the physical state of the absorbing gas and absorption spectrum. Both of the model types include hydrogen, helium, carbon, oxygen, silicon, and iron elements in the calculation. Many aspects of physics related to all the elements included are investigated thoroughly.

\item Our H II region models are able to reproduce the results of \cite{gar13}. These authors simulated a gas that included only hydrogen, and they considered only two energy levels and a plane-parallel approximation. Our results differ from theirs in our inclusion of  geometric dilution and the composition of the gas. The H II region model predicts that the gas in the ionized region evolves faster to equilibrium, whereas the gas beyond the IF evolves for a longer time ($\sim 10^{14}$ s). 

\item We have calculated  warm absorber spectra and their time evolution for a range of model properties. The models show different spectra depending on the flux and column density. Various absorption edges evolve on different timescales. Absorption edges evolve faster initially while slower at the later time when they approach the equilibrium, i.e., the optical depth of the gas changes faster near the  starting time and more slowly down later. More importantly, time dependent spectra do not look like the equilibrium spectra. This strengthens the need for time dependent calculation in the warm absorbers.

\item Density dependence affects the variability of the gas, and low-density gas evolves more slowly. Models with $n_H = 10^7$ cm$^{-3}$ approach equilibrium at $\sim 10^7$ s while the model with $10^{11}$ cm$^{-3}$ at $\sim 10^3$ s. Therefore, observations of time dependent absorption spectrum can be used to  constrain the  warm absorber density.

\item Time variability in the ionizing source and absorption light curve can be studied using Fourier transform. The Fourier transform shows a steeper light curve has more power at a lower frequency of variation of the ionizing source.  The warm absorber acts as a low-pass  filter applied to the power spectrum intrinsic to the source; this provides another way in which observations can be used to constrain the parameters of the warm absorber.
\end{itemize}

We thank Dr. Alan Hindmarsh from Lawrence Livermore National Laboratory for his help and valuable suggestions regarding the ordinary differential equation solver package. We are grateful to the anonymous referee for many constructive suggestions and comments. This work was supported by grants through the NASA Astrophysics Theory Program.

\newpage

\appendix
\section{Numerical approach}\label{if0}

The set of ordinary differential equations for our problem govern level populations, temperature, electron number density, and radiation field.  In the general case, this system of equations can have terms that cover a large range of values, owing to the diverse physical processes they describe.  This system is, therefore,  'stiff' \citep{coo69}; that is, they correspond to much faster changes in some variables than in others.  Some processes are important for small timescales, while others dominate for long timescales. 

If we suppose we have $r$ level population equations including all the elements included, one temperature equation, one electron fraction equation, and $s$ number of specific intensity equations, then the total number of equations would be $r+s+2$. This is for a single spatial point in the cloud. However, we have taken all the spatial points and corresponding equations at the same time and used a solver to go from one-time point to the next. If we take q number of spatial zone, we will have $(r+s+2)\times q$ number of equations. The nonlinear system of equations that arises at each time step after using backward Euler's method must be solved by a suitably powerful method, and that is some variation of Newton's method or secant method \citep{gal00}. These methods find the roots of the system of equations using iterative techniques. Newton's method for a system requires dealing with a large vector and the associated jacobian matrix to be solved for each time step. This method is expensive because it has to invert the large matrix multiple times until it meets the accuracy requirement. This is the reason we model the system to have a minimum number of equations to save computation time.

Equations (\ref{eqn5}), (\ref{eqn6}), (\ref{eqn7}) can be written as shown below.
\begin{eqnarray}
f_1(R_1,x_2,x_3.........,x_N)&=&0,\nonumber\\
f_2(R_1,x_2,x_3.........,x_N)&=&0,\nonumber\\
f_3(R_1,x_2,x_3.........,x_N)&=&0,\\
.\nonumber\\
.\nonumber\\
.\nonumber\\
f_N(R_1,x_2,x_3.........,x_N)&=&0\nonumber
\end{eqnarray}
Where $x_1=n_1^{t+1}, x_2=n_2^{t+1}$ and so on, where the superscript corresponds to the time step. All variables (level population, temperature, electron fraction and specific intensity) at time $t+1$ are replaced by $x_1, x_2 ....., x_N$ for the simplicity. The  goal is to find the roots of these equations. Suppose the above equations have the exact roots $(R_1^*, x_2^*, x_3^*,.......,x_N^*)$. Then we can write,
\begin{eqnarray}
f_1(R_1^*, x_2^*, x_3^*,.......,x_N^*)&=&0,\nonumber\\
f_2(R_1^*, x_2^*, x_3^*,.......,x_N^*)&=&0,\nonumber\\
f_3(R_1^*, x_2^*, x_3^*,.......,x_N^*)&=&0,\\
.\nonumber\\
.\nonumber\\
.\nonumber\\
f_N(R_1^*, x_2^*, x_3^*,.......,x_N^*)&=&0\nonumber
\end{eqnarray}

\noindent Using a Taylor series and neglecting the second and higher order terms \cite{new13} and writing in compact form, we have

\begin{equation}
    f_i(R_1^*, x_2^*, x_3^*,.......,x_N^*)=f_i(R_1, x_2, x_3,.......,x_N)+\sum_{j}(R_j^*-x_j) \frac{\partial f_i}{\partial x_j}+.......
\end{equation}

\noindent Where $x_1, x_2, x_3,.......,x_N$ are the initial guess of the variables.
We can write the above equation in vector notation as follows,

\begin{equation}\label{eqn8}
    \mathbf{f(R^*)=f(R) + J (R^*-x) + .......}
\end{equation}

\noindent Where $\mathbf{J}=\frac{\partial f_i}{\partial x_j}$ is the Jacobian matrix of the size $N \times N$. Since $\mathbf{x^*}$ is the exact root of the equations, we have $\mathbf{f(R^*)=0}$. Neglecting the higher order term, equation (\ref{eqn8})  becomes,
\begin{equation}\label{eqn9}
    \mathbf{J \Delta x =f(R)}
\end{equation}
where $\mathbf{\Delta x=x-x^*}$. The above equation is a system of linear equations of the form $\mathbf{Ax=v}$. This can be solved for $\mathbf{\Delta x}$ using various numerical methods.
Once we calculate $\mathbf{\Delta x}$, our new estimated vector will be $\mathbf{x^{'}=x-\Delta x}$ where $\mathbf{x^{*}}=\mathbf{x^{'}+\epsilon}$; $\epsilon$ is an error come from neglecting the higher order term in equation  (\ref{eqn8}).  Again, at this new set of values, we evaluate $\mathbf{f(R)}$ and calculate Jacobian matrix  and solve the equation (\ref{eqn9}) for $\mathbf{\Delta x}$. This will give $\mathbf{x^{''}=\mathbf{x'-\Delta x'}}$. We repeat this process until the required precision is met. In this way, we solve the above algebraic equations and find the values of the variables of interest from t to t+1. Once we find the solution at t+1, these values now act at the initial values, and we find another set of equations and solve at time t+2. This process continues until we reach the required final time point. In order to carry out this procedure, we implement the well-known differential equation solver, DVODE \citep{bro89}.

\section{Theoretical Calculation of Location of IF in H II region model}\label{if1}

An H II region corresponds to the region of ionized gas surrounding a source of ionizing radiation such as a star or white dwarf etc. So a source emitting a finite amount of ionizing radiation per second can only ionize the gas in a finite region. This yields a region where gas is fully ionized towards the source and nearly neutral on the rest of the cloud in spherical geometry. The region which separates these two ionized and neutral regions are called the ionization front (IF). A crude calculation of the IF location is obtained by setting the total number of ionization of an atom of a particular element in the gas per unit time equal to the total number of ionizing photons emitted by the source per unit time. In the case of equilibrium, the rate of ionization is equal to the rate of recombination in a volume V. Mathematically; the hydrogen ionization front can be calculated as \citep{ost89};

\begin{equation}\label{ifeqn1}
    \int_ {\varepsilon_{th}}^\infty \frac{L_\varepsilon d\varepsilon}{\varepsilon}= \int_ {0}^{r_1} n_e n_p \alpha_r(H^0,T) dV 
\end{equation}
where $L_\varepsilon$ is the specific luminosity of the ionizing source. In our model, we consider such a source of power $10^{32}$ erg s$^{-1}$. $n_e$ and $n_p$ are the number density of electrons and protons. $\alpha_r$ is the recombination rate of hydrogen ions. We consider the power-law continuum source of energy index of -1 that is $L_\varepsilon=L_0 \varepsilon^{-1}$. Since we have an inner radius and an outer radius in the case of the H II region model, the above equation (\ref{ifeqn1}) can be written as

\begin{equation}\label{ifeqn2}
    \int_ {\varepsilon_{th}}^\infty \frac{L_\varepsilon d\varepsilon}{\varepsilon}=\frac{4\pi}{3}(r_2^3-r_1^3) n_H^2 \alpha_r(H^0,T)
\end{equation}
where $r_1$ is the radius where IF exists and $r_2$ is the inner radius of the spherical gas cloud from the ionizing source. So for a simplistic approach, we assume the value of case A recombination coefficient of hydrogen ion $\alpha_r(H^0,T)\sim 10^{-13}$ cm$^3$ s$^{-1}$ \citep{ost89}. With this, the IF for initial ionizing luminosity ($10^{32}$ erg s$^{-1}$) exists at $\sim 10^{16}$ cm, and for changed luminosity ($3 \times 10^{32}$ erg s$^{-1}$), IF lies at $\sim 1.5 \times 10^{16}$ cm. In the same way, we can calculate the location of IF in the case of helium using the equation,

\begin{equation}\label{ifeqnhe}
    \int_ {\varepsilon_{1}}^\infty \frac{L_\varepsilon d\varepsilon}{\varepsilon}=\frac{4\pi}{3}(r_3^3-r_1^3) N_{He+}N_e\alpha_r(He^0,T)
\end{equation}
where $\varepsilon 1$ is the ionization potential of $He^0$, $r_3$ be the location of $He^+$ IF, $N_{He+}$ is the number density of singly ionized helium, $N_e=N_p+N_{He+}$ is the electron density, $\alpha_r(He^+,T)$ is the case A recombination coefficient of ionized helium. Using $10\%$ number density for helium and taking the recombination coefficient $\alpha_r(He^+,T)\sim 10^{-13}$ cm$^3$ s$^{-1}$ at around 25,000 K, the IF for $He^+$ to is found to be at $\sim 1.8 \times 10^{16}$ cm for initial flux and $\sim 2.7 \times 10^{16}$ cm for changed flux. These analytical results are compared with the numerical results in section 4.1. To express the IF quantitatively and for comparison, we define the IF as the region in the cloud where the ion fraction is greater than 0.5.
\newpage
\bibliography{tdp}{}
\bibliographystyle{aasjournal}
\end{document}